\documentclass[reprint,showpacs,preprintnumbers,amsmath,amssymb]{revtex4-1}

\pdfoutput=1
\usepackage[pdftex]{graphicx}
\usepackage{color}
\definecolor{darkred}{rgb}{0.4,0.0,0.0}
\definecolor{darkgreen}{rgb}{0.0,0.4,0.0}
\definecolor{darkblue}{rgb}{0.0,0.0,0.4}
\usepackage[bookmarks,linktocpage,colorlinks,linkcolor = darkred,urlcolor  = darkblue,citecolor = darkgreen,filecolor=cyano]{hyperref}
\usepackage{braket}
\usepackage[T1]{fontenc}
\usepackage{amsmath}
\usepackage{amsthm}
\usepackage{amssymb}
\usepackage{url}
\usepackage[utf8]{inputenc}
\usepackage[english]{babel}
\usepackage{dcolumn}
\usepackage{bm}
\usepackage{subfig}
\usepackage{float}
\usepackage{url} 
\usepackage{grffile}
\usepackage{xcolor}
\usepackage{booktabs}
\usepackage{verbatim}
\usepackage{graphicx}

\newcommand{\Tr}{\mbox{\rm Tr\,}}
\newcommand{\ReC}{\mbox{\rm Re\,}}

\newcommand{\be}{\begin{equation}}
\newcommand{\ee}{\end{equation}}
\newcommand{\bea}{\begin{eqnarray}}
\newcommand{\eea}{\end{eqnarray}}
\newcommand{\non}{\nonumber}
\newcommand{\bie}{\begin{small} \begin{itemize}}

\newcommand{\eie}{\end{itemize} \end{small}}
\newcommand{\LagFigSize}{0.95\columnwidth} 
\newcommand{\LLagFigSize}{1.05\columnwidth} 

\usepackage{cleveref}
\crefname{section}{Section}{Sections}
\Crefname{section}{Section}{Sections}
\crefname{equation}{Eq.}{Eqs.}
\Crefname{equation}{Eq.}{Eqs.}
\crefname{figure}{Fig.}{Figs.}
\Crefname{figure}{Fig.}{Figs.}
\crefname{table}{Table}{Tables}
\Crefname{table}{Table}{Tables}

\graphicspath{{./allplots/}}

\usepackage{ragged2e}

\begin{document}
%
\selectlanguage{english}
\title{%
Colour field densities of the quark-antiquark excited flux tubes in SU(3) lattice QCD
}
\author{P. Bicudo}
\email{bicudo@tecnico.ulisboa.pt}
\author{N. Cardoso}
\email{nuno.cardoso@tecnico.ulisboa.pt}
\author{M. Cardoso}
\email{marco.cardoso@tecnico.ulisboa.pt}
\affiliation{CeFEMA, Departamento de F\'{\i}sica, Instituto Superior T\'{e}cnico
(Universidade T\'{e}cnica de Lisboa),
Av. Rovisco Pais, 1049-001 Lisboa, Portugal}
\begin{abstract}
We combine techniques previously utilised to study flux tube field density profiles and to study the excited spectrum of the gluonic fields produced by a static quark-antiquark pair.
Working with pure gauge SU(3) fields discretised in a lattice, we utilise Wilson loops with a large basis of gluonic spacelike Wilson lines to include different excitations of the quark-antiquark flux tube.
To increase the signal over noise ratio, we use the multihit technique in the temporal Wilson lines and the APE smearing in spatial Wilson lines.
The number of gluonic operators combined with the space points where we compute the flux tube densities turns out to be very large, and we resort to GPUs and to CUDA codes. 
Computing the effective mass plot from the diagonalized correlation matrix,  we separate the excitations with different two-dimensional  angular momentum, parity and radial quantum numbers.
We then compute the colour field density profiles for all the components of the colour electric and colour magnetic fields. 
We analyse our results for the first excitations of the flux tube and search for signals of novel phenomena beyond the Nambu-Goto string model, such as a longitudinal mode or an explicit gluon.
\end{abstract}
\maketitle 

\section{Introduction}\label{intro}

Understanding the confinement of colour remains a main theoretical problem of modern physics. Its solution could also open the door to other unsolved theoretical problems. 
One of the evidences of confinement, where we may search for relevant details to understand it, is in the QCD flux tubes \cite{Wilson:1974sk}. 
Here we study quantitatively the excitations of the QCD flux tube with lattice QCD techniques, extending the work presented in Ref. \cite{Bicudo:2018yhk}.

Experimentally, we only have indirect evidence of the flux tubes, through the hadron spectrum and Regge trajectories 
\cite{Bicudo:2007wt}
who point to a linear confining potential 
\cite{Godfrey:1985xj,Isgur:1978xj}. 
A direct evidence for flux tube or gluonic excitations, would be the confirmed observation of hybrid mesons. A reliable quantitative lattice QCD prediction of flux tubes will assist our experimental colleagues in discovering these exotic mesons \cite{McNeile:2002az,McNeile:2006bz}, where the gluon degrees of freedom would excite quantum numbers inaccessible to the quark degrees of freedom. 

Presently, the qualitative understanding of flux tubes is quite developed, mostly based in string models.
The dominant behaviour of the flux tubes is clearly string-like, with a single scale $\sigma$. 
The main analytical model utilised in the literature to explain the behaviour of the QCD flux tubes is the Nambu-Goto bosonic string model \cite{Aharony:2009gg}. It assumes infinitely thin strings, with transverse quantum fluctuations only. The quantum fluctuations predict not only a zero mode width of the groundstate flux tube, increasing with distance \cite{Gliozzi:2010zv}, but also an infinite tower of quantum excitations \cite{Alvarez:1981kc,Arvis:1983fp}. 
Both effects have been observed by lattice QCD computations \cite{Campbell:1987nv,Perantonis:1990dy,Lacock:1996ny,Lacock:1996vy,Juge:1999ie,Juge:2002br}, indeed confirming the string dominance of the QCD flux tube.
The string-like behaviour partly obscures the details of confinement or of other possible hadronic phenomena, and precise lattice QCD computations are necessary to go beyond the string models.

Recently, our lattice QCD collaboration  PtQCD 
\cite{PtQCD} 
studied the zero temperature groundstate flux tube of pure gauge QCD, and found evidence for a penetration length $\lambda$ 
\cite{Cardoso:2013lla}, 
as a second scale other than the string tension $\sigma$, contributing to the colour fields density profile of the flux tube. 

Another instance where the flux tube deviates from the string model is at short quark-antiquark distances, where the fields produced by the charges diverge, and where lattice QCD has shown the potential becomes dominated by perturbative QCD 
\cite{Karbstein:2014bsa}.

A lattice QCD evidence for explicit gluon degrees of freedom or for longitudinal quantum excitations would also go beyond the Nambu-Goto string model.

in Section \ref{sec:framework} we thus combine different lattice QCD techniques adequate to study flux tube field density profiles  \cite{Cardoso:2013lla,Mueller:2018fkg} and to study the excited spectrum of the gluonic fields \cite{Campbell:1987nv,Perantonis:1990dy,Lacock:1996ny,Lacock:1996vy,Juge:1999ie,Juge:2002br} produced by a static quark-antiquark pair.
Working with pure gauge SU(3) fields discretised in a lattice, we utilise Wilson loops with a large basis of gluonic spacelike Wilson lines to include different excitations of the quark-antiquark flux tube. We combine our operators, to block diagonalize our basis the angular momentum and parity quantum numbers of the $D_{\infty h}$ point group.
We numerically diagonalise the remaining blocks of the correlation matrix and compute the corresponding effective mass plots.
We then compute the field density profiles for all the components of the colour electric and colour magnetic fields. 
We also discuss our computational efficiency. The number of gluonic operators combined with the space points where we compute the flux tube densities turn out to be very large, and we resort to computations in GPUs and to CUDA codes. 

In Section \ref{sec:results} we show, for the quantum numbers where the signal is clear, the results of our computations for the spectra and the field densities of our flux tubes.
Finally in Section \ref{sec:analysis}, we analyse our results for the first excitations of the flux tube and search for signals of novel phenomena beyond the Nambu-Goto string model; in Section \ref{sec:conclu} we conclude our work.

\section{Lattice QCD framework to compute the flux tubes \label{sec:framework}}
\subsection{Our 33 operator basis to produce the different excited quantum numbers \label{sec:operator}}

\begin{figure}[t!]
\begin{centering}
\includegraphics[width=0.95\columnwidth]{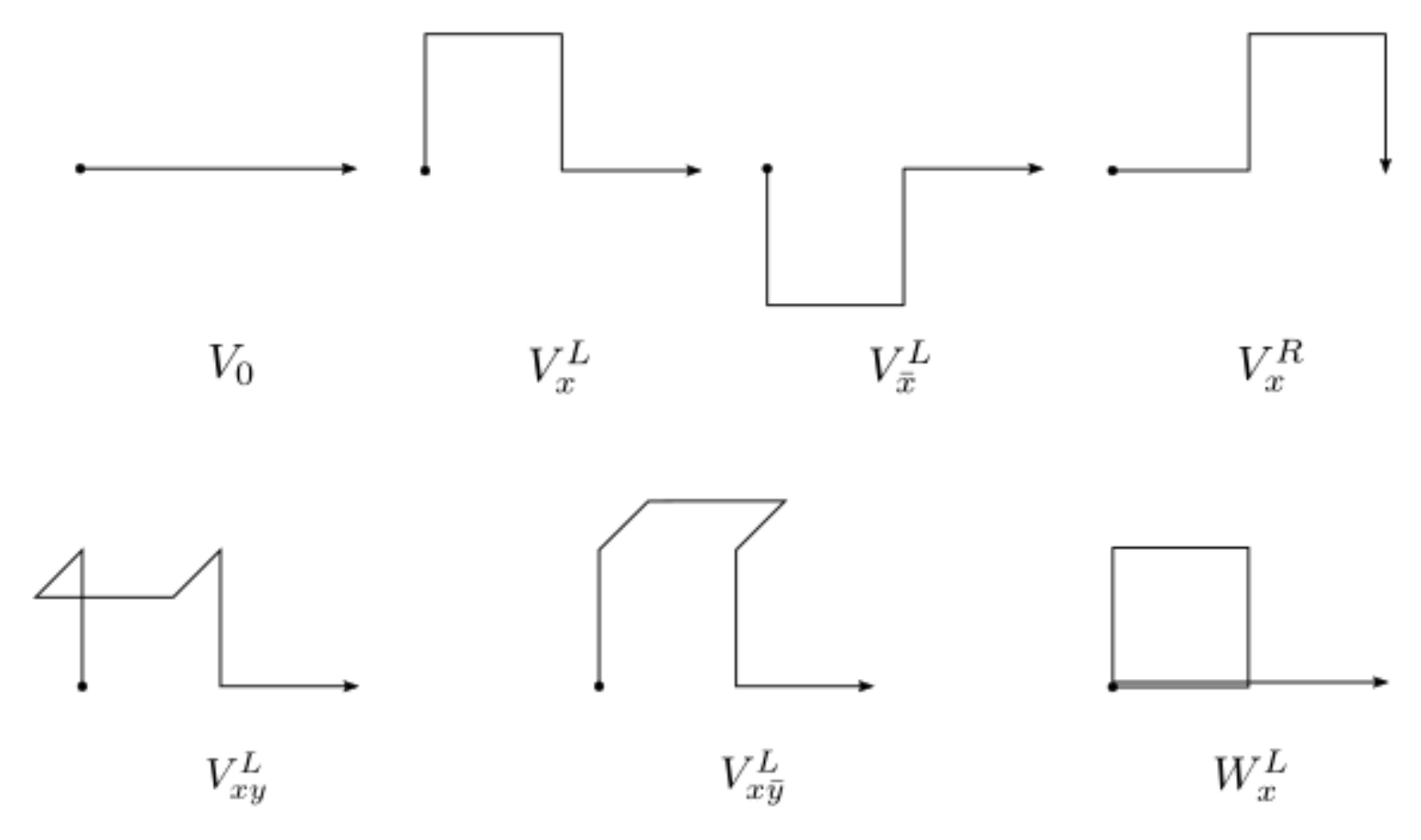}\caption{Examples of the paths from the quark to the antiquark used to construct the gauge field operators.
\label{fig:basisex}}
\par\end{centering}
\end{figure}

In the study of the flux tubes, we utilise a basis of spacial Wilson line operators, defined in Fig.  \ref{fig:basisex}, sufficiently complete to include different types of flux tube excitations \cite{Lacock:1996vy}. Since we have static charges, our temporal Wilson lines are straight, and they close the Wilson loop.
As usual we choose our frame such that the charge axis is the $z$ axis, and the origin is set at the midpoint between the quark and the antiquark, with distance $R$. The $x$ and $y$ axis are in the two perpendicular directions.
 
Our basis is composed by four kinds of operators.
\begin{itemize}
\item The direct operator $V_{0}$.
\item The eight open-staple operators $V_{x}^{L}$ , $V_{y}^{L}$, $V_{\bar{x}}^{L}$,
$V_{\bar{y}}^{L}$, $V_{x}^{R}$, $V_{y}^{R}$, $V_{\bar{x}}^{R}$
and $V_{\bar{y}}^{R}$.
\item The sixteen open-staple two-direction operators $V_{xy}^{L}$, $V_{x\bar{y}}^{L}$,
$V_{\bar{x}y}^{L}$, $V_{\bar{x}\bar{y}}^{L}$, $V_{yx}^{L}$, $V_{y\bar{x}}^{L}$,
$V_{\bar{y}x}^{L}$, $V_{\bar{y}\bar{x}}^{L}$, $V_{xy}^{R}$, $V_{x\bar{y}}^{R}$,
$V_{\bar{x}y}^{R}$, $V_{\bar{x}\bar{y}}^{R}$, $V_{yx}^{R}$, $V_{y\bar{x}}^{R}$,
$V_{\bar{y}x}^{R}$ and $V_{\bar{y}\bar{x}}^{R}$.
\item The eight closed-staple operators similar to the open-staple ones
$W_{x}^{L}$, $W_{y}^{L}$, $W_{\bar{x}}^{L}$,
$W_{\bar{y}}^{L}$, $W_{x}^{R}$, $W_{y}^{R}$, $W_{\bar{x}}^{R}$ and $W_{\bar{y}}^{R}$.
\end{itemize}
The bar over a coordinate index means that there is displacement in the negative axis direction.
The $L$ (side of the static antiquark) and $R$ (side of the static quark) labels indicate whether the staple is in the left
or in the right of the origin.

%

All our staples in the operators are long, they are implanted in one half of the Wilson line, with length $R/2$. We opt for long operators because we want them to represent the first excitations of the flux tube. While $V_0$ has the minimum number of links, the one direction staple operator has two more links, the two direction operator has four more links and the closed-staple operator $W$ has two plus $R/a$ more links.

This amounts to 33 different operators. Since the computation of the flux tube profiles is extremely demanding, although it would be interesting to use a more complete basis with more operators, we limit our basis to the present 33 operators. 

To further limit the size of the correlation matrix, we first block diagonalise it.
With linear combinations of our operators, we construct operators with a definite symmetry, since operators in different representations do no mix.

%
\begin{figure}[t!]
\begin{centering}
\includegraphics[width=1.05\columnwidth]{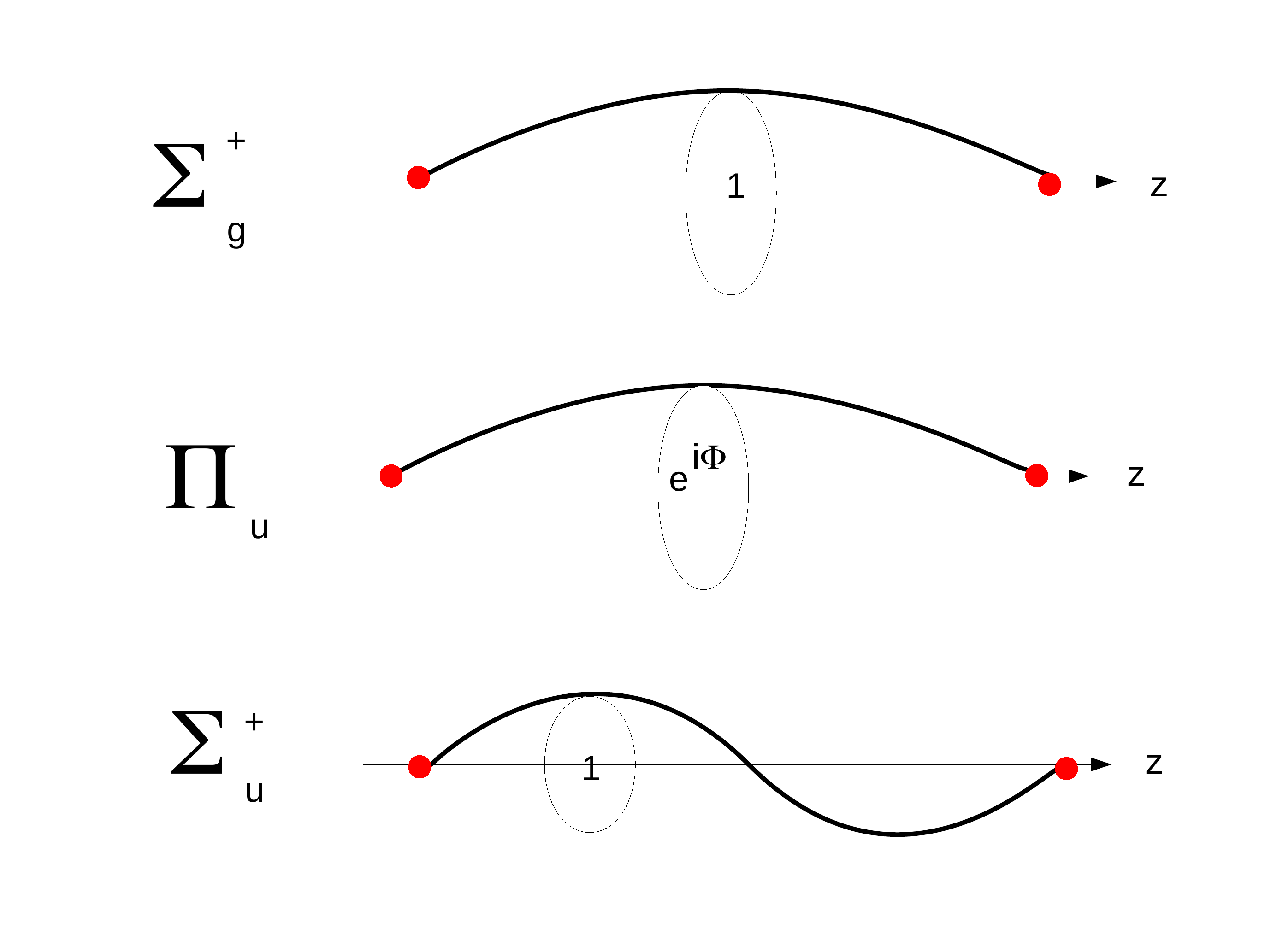}
\par\end{centering}
\caption{Sketches of the simplest quantum numbers we can excite in a flux tube, labelled by the point group representations of a homonuclear diatomic molecule.
\label{fig:quantumnumbers}}
\end{figure}

The symmetry group of our flux tubes, with two static sources, is equivalent to the one of the molecular orbitals of homonuclear diatomic molecules. It is the point group denominated $D_{\infty h}$. We thus utilise the standard quantum number notation of molecular physics, already adopted in the previous studies of QCD flux tube excitations \cite{Campbell:1987nv,Perantonis:1990dy,Lacock:1996ny,Lacock:1996vy,Juge:1999ie,Juge:2002br},  
see also the recent Ref. \cite{Reisinger:2017btr}. 
$D_{\infty h}$ has three symmetry sub-groups, and they determine three quantum numbers. 

\begin{itemize}
\item{\bf Two-dimensional rotation about the charge axis}
\\
The two-dimensional rotation about the charge axis corresponds to the quantum angular number, projected in the unit vector of the charge axis  $\Lambda = \left| {\bf J}_g\!\cdot \hat e_z \right| $. The capital Greek
letters $\Sigma, \Pi, \Delta, \Phi, \dots$ indicate as usually states
with $\Lambda=0,1,2,3,\dots$, respectively. The notation is reminiscent of the $s, \, p, \, d \cdots $ waves in atomic physics.  In the case of two-dimensional rotations there are only two projections ${\bf J}_g\!\cdot \hat e_z = \pm \Lambda$.

\item{\bf Parity inversion about the median point} 
\\
The permutation of the quark and the antiquark static charges is equivalent to a combined operations of
charge conjugation and spatial inversion about the origin. Its eigenvalue is denoted by
$\eta_{CP}$.  States with $\eta_{CP}=1 (-1)$ are denoted
by the subscripts $g$ ($u$), short notation for {\em gerade} ( {\em ungerade}).  

\item{\bf Additional parity}
\\
Moreover there is a third quantum number, different from the phase corresponding to a two-dimensional p-wave. 
Due to the planar, and not three-dimensional, angular momentum there is an additional label for the s-wave
$\Sigma$ states only. $\Sigma$ states which
are even (odd) under the reflection about a plane containing the molecular
axis are denoted by a superscript $+$ $(-)$. 

\end{itemize}

With these quantum numbers, the energy levels of the flux tubes are labeled  as $\Sigma_g^+$, $\Sigma_g^-$, $\Sigma_u^+$, $\Sigma_u^-$,
$\Pi_u$, $\Pi_g$, $\Delta_g$, $\Delta_u \cdots$
We illustrate the simplest quantum numbers in Fig. \ref{fig:quantumnumbers}.
As a result of the different symmetries and respective quantum numbers, we rearrange our initial 33 operators into the following operators.

\begin{widetext}

For the groundstate quantum numbers ${\Sigma_{g}^{+}}$, we have four operators:
\begin{eqnarray}
\mathcal{A}_{0,1} & = & V_{0}
\non
\\
\mathcal{A}_{0,2} & = & \frac{1}{2\sqrt{2}}\big(V_{x}^{L}+V_{y}^{L}+V_{\bar{x}}^{L}+V_{\bar{y}}^{L}+V_{x}^{R}+V_{y}^{R}+V_{\bar{x}}^{R}+V_{\bar{y}}^{R}\big)
\non
\\
\mathcal{A}_{0,3} & = & \frac{1}{4}\big(V_{xy}^{L}+V_{x\bar{y}}^{L}+V_{\bar{x}y}^{L}+V_{\bar{x}\bar{y}}^{L}+V_{yx}^{L}+V_{y\bar{x}}^{L}+V_{\bar{y}x}^{L}+V_{\bar{y}\bar{x}}^{L}
 \ +V_{xy}^{R}+V_{x\bar{y}}^{R}+V_{\bar{x}y}^{R}+V_{\bar{x}\bar{y}}^{R}+V_{yx}^{R}+V_{y\bar{x}}^{R}+V_{\bar{y}x}^{R}+V_{\bar{y}\bar{x}}^{R}\big)
 \non
 \\
\mathcal{A}_{0,4} & = & \frac{1}{2\sqrt{2}}\big(W_{x}^{L}+W_{y}^{L}+W_{\bar{x}}^{L}+W_{\bar{y}}^{L}+W_{x}^{R}+W_{y}^{R}+W_{\bar{x}}^{R}+W_{\bar{y}}^{R}\big)
\end{eqnarray}

We have four operators for the quantum numbers ${\Pi_{u}}$, with projection $ {\bf J}_g\!\cdot \hat e_z=+1$:
\begin{eqnarray}
\mathcal{A}_{4,1} & = & \frac{1}{2\sqrt{2}}\big(V_{x}^{L}+iV_{y}^{L}-V_{\bar{x}}^{L}-iV_{\bar{y}}^{L}+V_{x}^{R}+iV_{y}^{R}-V_{\bar{x}}^{R}-iV_{\bar{y}}^{R}\big)
\non
\\
\mathcal{A}_{4,2} & = & \frac{1}{4}\big(V_{xy}^{L}+V_{x\bar{y}}^{L}-V_{\bar{x}y}^{L}-V_{\bar{x}\bar{y}}^{L}+iV_{yx}^{L}+iV_{y\bar{x}}^{L}-iV_{\bar{y}x}^{L}-iV_{\bar{y}\bar{x}}^{L} 
+V_{xy}^{R}+V_{x\bar{y}}^{R}-V_{\bar{x}y}^{R}-V_{\bar{x}\bar{y}}^{R}+iV_{yx}^{R}+iV_{y\bar{x}}^{R}-iV_{\bar{y}x}^{R}-iV_{\bar{y}\bar{x}}^{R}\big) 
\non
\\
\mathcal{A}_{4,3} & = & \frac{1}{4}\big(V_{xy}^{L}-V_{x\bar{y}}^{L}+V_{\bar{x}y}^{L}-V_{\bar{x}\bar{y}}^{L}-iV_{yx}^{L}+iV_{y\bar{x}}^{L}-iV_{\bar{y}x}^{L}+iV_{\bar{y}\bar{x}}^{L} 
+V_{xy}^{R}-V_{x\bar{y}}^{R}+V_{\bar{x}y}^{R}-V_{\bar{x}\bar{y}}^{R}-iV_{yx}^{R}+iV_{y\bar{x}}^{R}-iV_{\bar{y}x}^{R}+iV_{\bar{y}\bar{x}}^{R}\big)
\non
\\
\mathcal{A}_{4,4} & = & \frac{1}{2\sqrt{2}}\big(W_{x}^{L}+iW_{y}^{L}-W_{\bar{x}}^{L}-iW_{\bar{y}}^{L}+W_{x}^{R}+iW_{y}^{R}-W_{\bar{x}}^{R}-iW_{\bar{y}}^{R}\big)
\end{eqnarray}

For the quantum numbers  ${\Sigma_{u}^{+}}$, we have three operators:
\begin{eqnarray}
\mathcal{A}_{2,1} & = & \frac{1}{2\sqrt{2}}\big(V_{x}^{L}+V_{y}^{L}+V_{\bar{x}}^{L}+V_{\bar{y}}^{L}-(V_{x}^{R}+V_{y}^{R}+V_{\bar{x}}^{R}+V_{\bar{y}}^{R})\big)
\non
\\
\mathcal{A}_{2,2} & = & \frac{1}{4}\big(V_{xy}^{L}+V_{x\bar{y}}^{L}+V_{\bar{x}y}^{L}+V_{\bar{x}\bar{y}}^{L}+V_{yx}^{L}+V_{y\bar{x}}^{L}+V_{\bar{y}x}^{L}+V_{\bar{y}\bar{x}}^{L}
\ -(V_{xy}^{R}+V_{x\bar{y}}^{R}+V_{\bar{x}y}^{R}+V_{\bar{x}\bar{y}}^{R}+V_{yx}^{R}+V_{y\bar{x}}^{R}+V_{\bar{y}x}^{R}+V_{\bar{y}\bar{x}}^{R})\big)
\non
\\
\mathcal{A}_{2,3} & = & \frac{1}{2\sqrt{2}}\big(W_{x}^{L}+W_{y}^{L}+W_{\bar{x}}^{L}+W_{\bar{y}}^{L}-(W_{x}^{R}+W_{y}^{R}+W_{\bar{x}}^{R}+W_{\bar{y}}^{R})\big)
\end{eqnarray}

\end{widetext}

These are the states with lowest energy, $\Sigma^+_g$, $\Pi_u$ and $\Sigma^+_u$, illustrated in Fig. \ref{fig:quantumnumbers}. In what concerns the combinations of operators with the remaining quantum numbers  $\Sigma_{g}^{-}, \ \Sigma_{u}^{-}, \ \Pi_{g}, \Delta_g, \ \Delta_u \cdots $, we also studied them. Due to their higher complexity and  energy and we did not get a clear enough signal for the respective flux tube. Thus we do not find relevant to list their respective combination of operators here. A larger basis of operators, more configurations and more effective noise reduction techniques will be necessary to study them. However they require computational power beyond our resources and we leave this for the future.

\subsection{Computation of the excited state spectra \label{sec:spectra}}

We start by utilizing the correlation matrix $\langle {\mathcal W}_{kl}(t) \rangle$ to compute the energy levels of the excited states, as done previously in the literature. 
Now the sub-indices $k$ and $l$ stand for the spacial operators in the operator basis defined in Section  \ref{sec:operator}, denoted $O_k$.
The spacial operators are connected by temporal Wilson lines $L$,
\begin{eqnarray}
 {\mathcal W}_{kl}(t) &=&
 O_k(-\mathbf R/2, \mathbf R/2,-t/2) \, 
 L(\mathbf R /2, -t/2, t/2) \, 
\\
 \non
 &&
 O_l^\dagger(\mathbf R/2, - \mathbf R /2, t/2) \,   
 L^\dagger(- \mathbf R /2, t/2, -t/2) \ .
\end{eqnarray}
The statistical average $\langle \cdots \rangle$ is performed over our ensemble of gauge link configurations.

Notice each matrix element corresponds to an evolution operator in Euclidean space, where all energy levels $E_i$ contribute, with coefficients depending on how close the operator is to the actual physical states, with the  Euclidean damping factor $\exp( -E_i \, t)$.

The first step to compute the energy levels, is to diagonalise the correlation matrix 
$\langle {\mathcal W}_{kl}(t) \rangle$, for each time extent $t$ of the Wilson loop, and get a set of time dependent eigenvalues $\lambda_i(t)$.
With the time dependence, we study the effective mass plot
\begin{equation}
E_i \simeq - \log { \lambda_i(t+1) \over \lambda_i(t)} \ ,
\end{equation}
and search for clear plateaux consistent with a constant energy $E_i$ in intervals $ t \in [{t_i}_\text{ini}, {t_i}_\text{fin}]$ between the initial and final time of the plateau.
The different energies  levels $E_i$, should correspond to the groundstate and excited states of the flux tube. If our operator basis is good enough, then $E_0$ is extremely close to the groundstate energy, $E_1$ is very close to the the first excited state, etc.

%
\begin{figure}[t!]
\begin{centering}
\includegraphics[width=1.00\columnwidth]{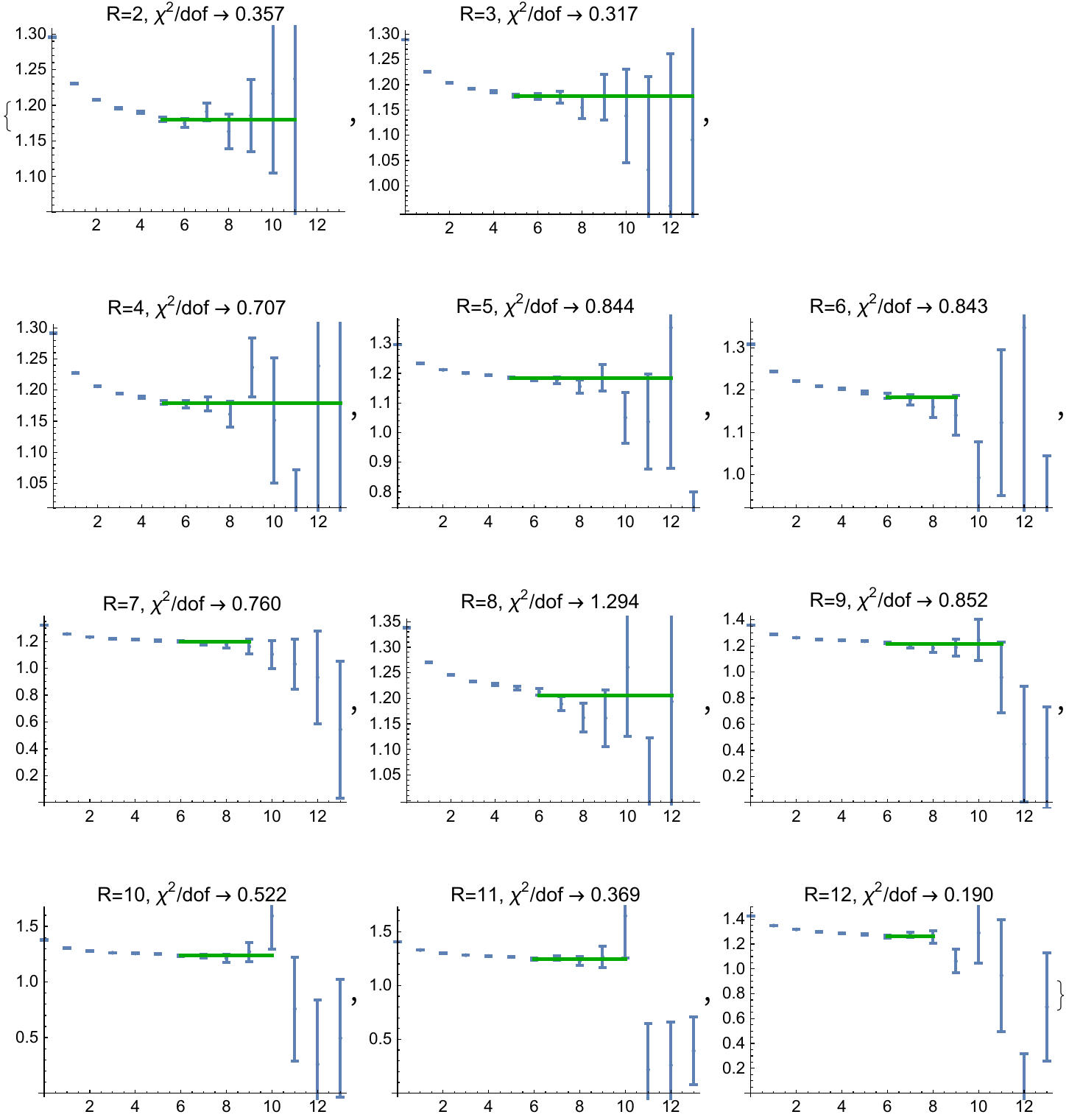}
\par\end{centering}
\caption{Effective mass plots for the potential V(R) of the ${\Sigma_g^+}^*$ excited state. The potential and distance are in units of lattice spacing $a$.
\label{fig:emp}}
\end{figure}

Moreover, with the diagonalisation 
we also obtain the eigenvector operators corresponding to the groundstate, first excitation, etc. 
We get a linear combination of our initial operators,
\begin{eqnarray}
\label{eq:eigen}
\widetilde O_0=c_{01}\, O_1+ c_{02} \, O_2 + \cdots
\\
\nonumber
\widetilde O_1=c_{11} \, O_1+ c_{12} \, O_2 + \cdots
\\
\nonumber
\cdots
\end{eqnarray}
Notice this result must be interpreted with a grain of salt. 
The eigenvector operators $\widetilde O_i$ do not exactly correspond to the respective state as in quantum mechanics,
but they get the clearest possible signal to noise ratio, among our operator basis.

The eigenvector operators $\widetilde O_i$ and the respective correlation matrix can be used in the same time interval $ t \in [{t_i}_\text{ini}, {t_i}_\text{fin}]$ ideal for the effective mass plateaux of the energy spectrum.
In Fig. \ref{fig:emp} and Table \ref{tab:emp}, we show the effective mass plots, in the case of  the excited state ${\Sigma^+_g}^*$ (eigenvalue 2 of our correlation matrix).

\begin{table}[t!]
\begin{center}
\begin{tabular}{c|ccc|cc}
\hline
R & V(R) & err & $\chi^2$/dof & ${t_2}_\text{ini}$ & ${t_2}_\text{fin}$ \\
\hline
2 & 1.17984 & 0.00236283 & 0.35667 & 5 & 11 \\
3 & 1.17755 & 0.00202011 & 0.316707 & 5 & 13 \\
4 & 1.17899 & 0.00213264 & 0.707024 & 5 & 13 \\
5 & 1.18287 & 0.00474316 & 0.843876 & 5 & 12 \\
6 & 1.1832 & 0.00929156 & 0.843143 & 6 & 9 \\
7 & 1.19658 & 0.0114933 & 0.760257 & 6 & 9 \\
8 & 1.20536 & 0.0207695 & 1.29409 & 6 & 12 \\
9 & 1.21672 & 0.0188055 & 0.85246 & 6 & 11 \\
10 & 1.23615 & 0.00494562 & 0.522346 & 6 & 10 \\
11 & 1.24705 & 0.00467709 & 0.369248 & 6 & 10 \\
12 & 1.26087 & 0.00771923 & 0.189831 & 6 & 8 \\
\hline
\end{tabular}
\caption{\label{tab:emp} Computing the potential  V(R) of the ${\Sigma_g^+}^*$ excited state (eigenvalue 2 of our correlation matrix) with the effective mass plots. The potential and distance are in units of lattice spacing $a$.}
\end{center}
\end{table}

\subsection{Computation of the chromofields in the flux tube \label{sec:crhomofields}}

We start by reviewing the technique of Ref. \cite{Cardoso:2013lla} since we utilise it to compute the chromomagnetic fileds. Let us temporarily assume we have a simple quark-antiquark Wilson loop $\cal W$. As in Ref. \cite{Cardoso:2013lla}, the central observables that govern the event in the flux tube can be extracted from the correlation of a plaquette $\square_{\mu\nu}$ with the Wilson loop $\cal W$,
\begin{equation}
f_{\mu\nu}(R,r) = \frac{\beta}{a^4} \left[\frac{\Braket{\mathcal{W}(\mathbf R,t)\,\square_{\mu\nu}(\mathbf r)}}{\Braket{\mathcal{W}(\mathbf R,t)}}-\Braket{\square_{\mu\nu}(\mathbf r)}\right]
\label{eq:field}
\end{equation}
where $\mathbf r=(x,y,z)$ denotes the spacial distance of the plaquette from the centre of the line segment connecting the quark sources, $R$ is the quark-antiquark separation and $t$ is the time extent of the Wilson loop.
Our plaquette is defined as,
\bea
\square_{\mu\nu}\Bigl(\mathbf r  +{ \mu + \nu \over 2} \Bigr) &=&1 - \frac{1}{3} \ReC\,\Tr \Bigl[ U_{\mu}(\mathbf r) U_{\nu}(\mathbf r+\mu) 
\non
\\
&& \times U_{\mu}^\dagger(\mathbf r+\nu) U_{\nu}^\dagger(\mathbf r) \Bigr]\ .
\label{eq:plaquettewithtrace}
\eea
Expanding it in powers of the small lattice spacing $a$ we get,
\bea
\square_{\mu\nu}  &=&  1 - {1 \over 3} \ReC \,  \Tr \exp  \left[ i  g a^2 \sum_c   F_{\mu\nu} ^c  T^c + {\cal O} (a^3) \right]
\non \\ 
&=&  {a^4 \over 2 \beta}  \left[ \sum_c F_{\mu\nu} ^c  \, F_{\mu\nu} ^c   + 
 {\cal O}(a) 
\right] \ .
\label{notrace}
\eea
Notice in non-Abelian gauge theories, such as SU(3), the electric and magnetic field components are not gauge invariant since they depend on the colour index $c$. 
We have to go up to order $a^4$ to find our first non-vanishing gauge invariant term in the plaquette expansion, and it is the square of a component of the electric or magnetic fields. For instance ${E_x}^2= \sum_c ( {E_x}^c)^2 $ is gauge invariant, while ${E_x}^c$ is not.

Therefore, using all the different plaquette orientations $(\mu,\nu)=(2,3), (3,1), (1,2),$ $(1,4),(2,4),(3,4)$, we can respectively relate the six components in \cref{eq:field} to the components of the chromoelectric and chromomagnetic fields,
\begin{equation}
f_{\mu\nu}\rightarrow\left(\Braket{B_x^2},\Braket{B_y^2},\Braket{B_z^2},\Braket{E_x^2},\Braket{E_y^2},\Braket{E_z^2}\right) \ .
\end{equation}
Notice these are the Euclidan space components. In Minkowski space we must include a $-$ phase in the magnetic field density, $B_i^2 \to - B_i^2$. 
With the field densities it is then trivial to compute  the total action (Lagrangian) density, $\Braket{\mathcal{L}}=\frac12\left(\Braket{E^2}-\Braket{B^2}\right)$.

%
\begin{figure}[t!]
\begin{centering}
\includegraphics[width=1.00\columnwidth]{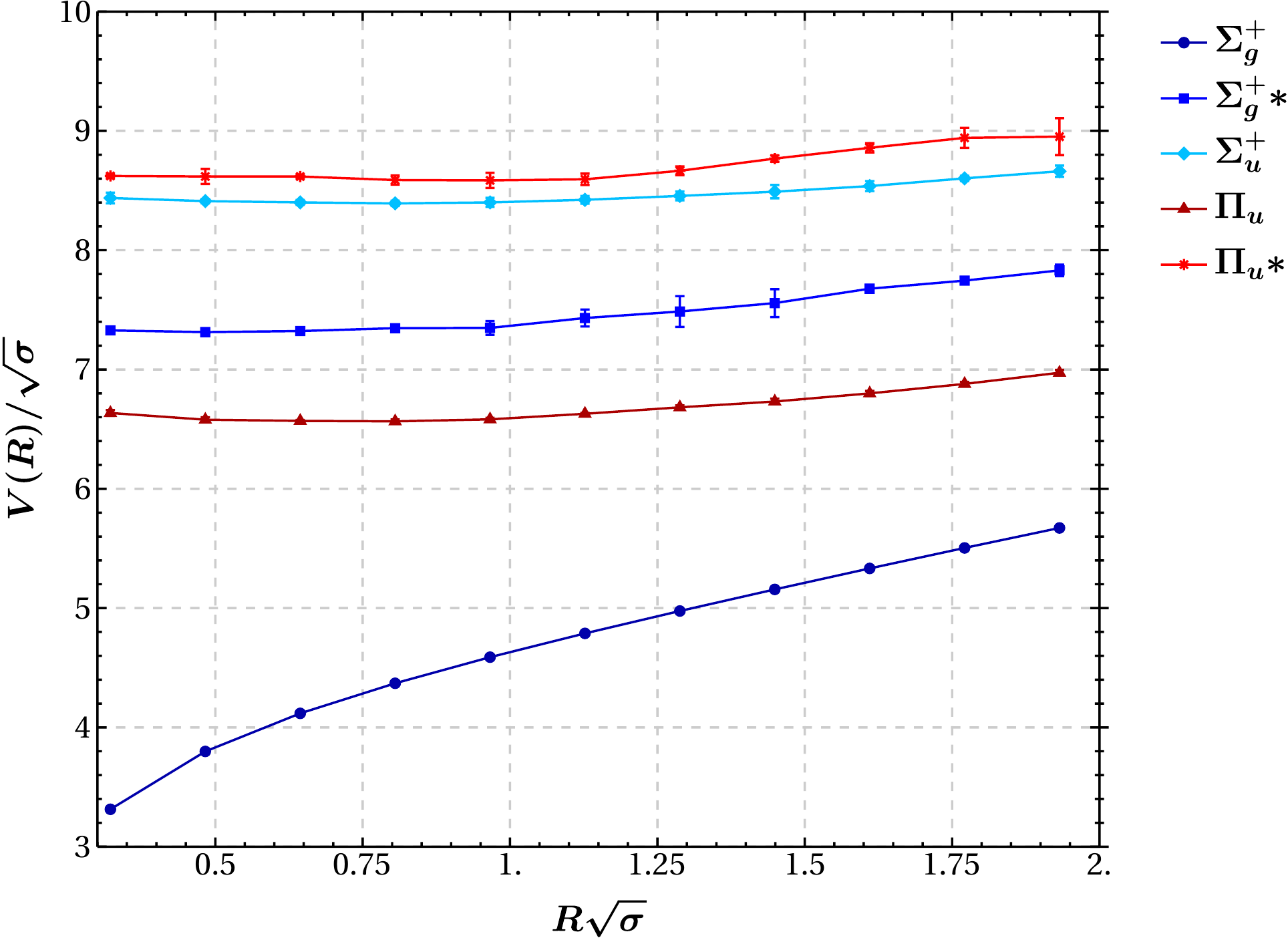}
\par\end{centering}
\caption{Flux tube spectra $V(R)$  as a function of the charge distance $R$. The distance and energy are shown in string tension units $\sqrt \sigma$. We only show the spectra of the quantum numbers $\Sigma^+_g, \Pi_u$ and $\Sigma^+_u$, producing the clearest signals for the flux tube. 
\label{fig:pot}}
\end{figure}

Now, to extend Eq. (\ref{eq:field}) for the study of excited flux tubes, we simply have to replace the Wilson loop $\cal W$ by $ \widetilde {\cal{W}}_i$, where the spacial links are given by the eigenvector operators $\widetilde O_i$ of Eq. (\ref{eq:eigen}).

The eigenvector operators $\widetilde O_i$ and the respective Wilson loop $ \widetilde {\cal{W}}_i$ can be used in the same time interval $ t \in [t_\text{ini}, t_\text{fin}]$ ideal for the effective mass plateaux of the energy spectrum.
	
\subsection{Configuration ensemble and code efficiency \label{sec:efficiency}}

We compute our results using 1199 configurations for a fixed lattice volume of $24^3\times 48$ and $\beta=6.2$. 
Our figures are presented in lattice spacing units of $a$, with $a = 0.07261(85)\, \text{fm}$ or $a^{-1} = 2718(32)\, \text{MeV}$.
The quark and antiquark are located at $(0, 0, -R/2)$ and $(0, 0, R/2)$ for $R$ between 6 and 10 in lattice spacing units.

Moreover, in order to improve the signal over noise ratio, we use the multihit technique  in the temporal Wilson lines and the APE smearing spatial Wilson lines \cite{Cardoso:2013lla}.
The multihit technique,  \cite{Brower:1981vt, Parisi:1983hm}, replaces each temporal link by its thermal average,
\begin{equation}
	U_4\rightarrow \bar{U}_4=\frac{\int dU_4 U_4 \,e^{\beta\Tr \left[U_4 F^\dagger\right]}}{\int dU_4 \,e^{\beta\Tr\left[ U_4 F^\dagger\right]}} \ .
\end{equation}
Here it is not possible to utilise the extended multihit technique as defined in Ref.  \cite{Cardoso:2013lla}, because our operators in the spatial Wilson line have a broader structure.

%
\begin{figure}[t!]
\begin{centering}
\includegraphics[width=1.00\columnwidth]{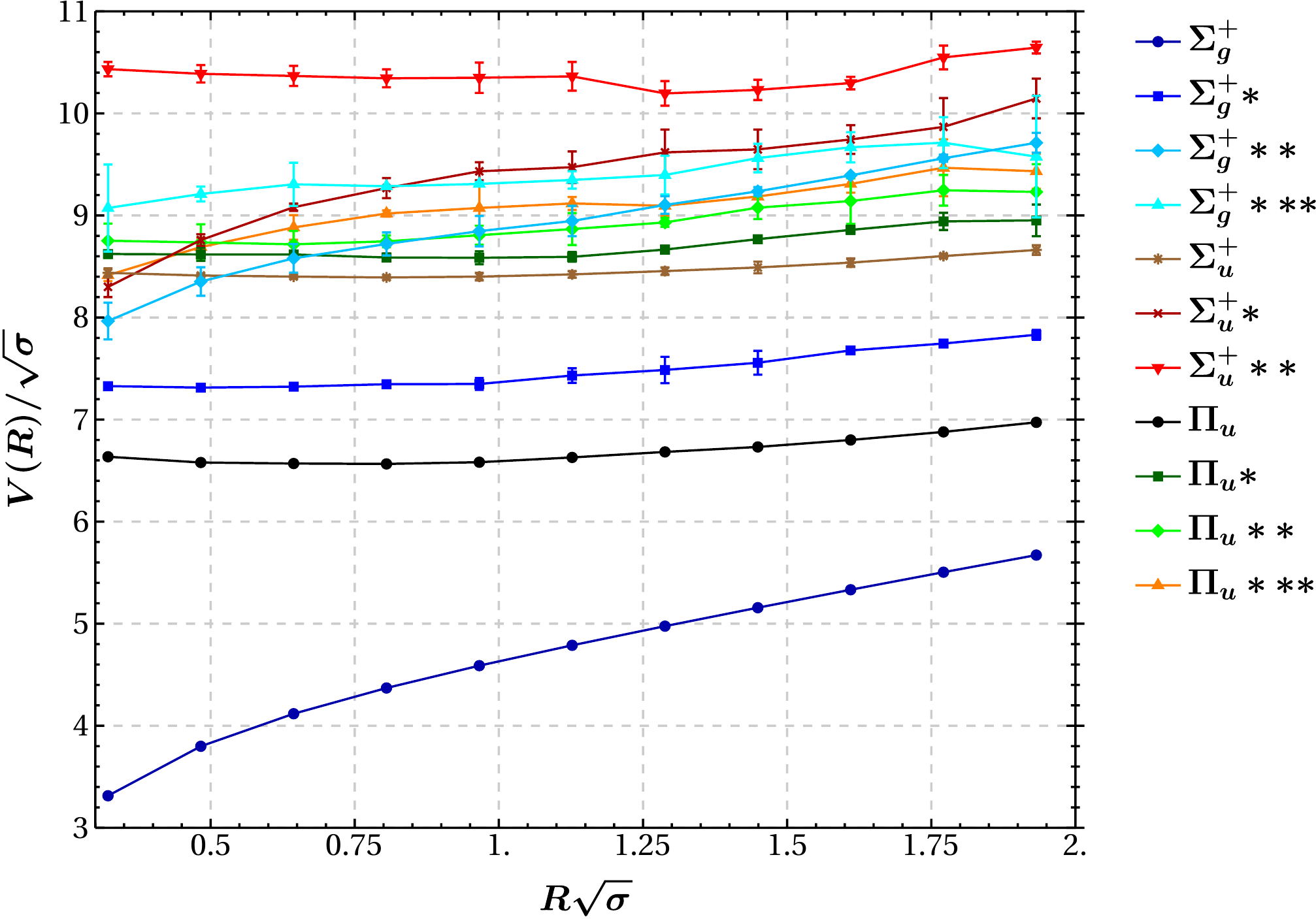}
\par\end{centering}
\caption{Flux tube spectra $V(R)$  as a function of the charge distance $R$. 
We now show all the spectra with clear plateaux in the effective mass plots, including the spectra of Fig. \ref{fig:pot} and the levels we discard.
\label{fig:pot_all}}
\end{figure}

The number of gluonic operators combined with the number space points where we compute the flux tube densities turns out to be very large, requiring a large computer power. We thus write all our codes in CUDA and run them in computer servers with NVIDIA GPUs. The computation of the chromofields are very computer intensive and due to the GPU limited memory this requires an intensive use of atomic memory operations. 

Moreover we simplify the possible number of operators. Compared with Ref. \cite{Juge:1999ie,Juge:2002br} who specialized in computing the spectrum, we have to utilise a smaller set of operators. We also limit the number of inter-charge distances, we compute the fields for inter-charge distances of $R=6 a, \ 8 a, \ 10 a$. In string tension units, we have  $a= 0.161013 / \sqrt \sigma$, for instance $6a= 0.966077 / \sqrt \sigma$.

For example, to calculate the field densities, per configuration and per flux tube state, our CUDA code takes approximately 70 min to run on a 
GeForce GTX TITAN 3.5cc (architecture Kepler)
and approximately 50 on a
GeForce GTX TITAN X 5.2cc (architecture Maxwell).

\section{Results \label{sec:results}}

In Fig. \ref{fig:pot} we show our results for the flux tube spectra, as a function of the charge distance $R$. The distance and energy are shown in string tension lattice unit $\sqrt \sigma $.

The ground-state $\Sigma^+_g$ is the familiar static-quark potential \cite{Cardoso:2013lla}. The lowest-lying excitation is the $\Pi_u$, it has two-dimensional angular momentum ${\bf J}_g\!\cdot \hat e_z =\pm 1$. 
Then the next excitation is the first radial excitation of the fundamental state, ${\Sigma^+_g}^*$.
The only other quantum number with clear results for the flux tube is the s-wave with inversion parity, corresponding to the first excited harmonic  $\Sigma^+_u$, as shown in Fig. \ref{fig:quantumnumbers}.
For all remaining quantum numbers our flux tubes have larger error bars, and with our present computational resources we abandon the pursue of their study.

For these quantum numbers, we get as many excited levels with clear plateaux in the effective mass plots as the number of operators we have. However, we should trust less states than the ones we observe clearly. We only accept a number of states smaller than half of the respective number of operators, and the respective spectrum is shown in Fig. \ref{fig:pot}. 
For instance in quantum mechanics, when using a limited basis of states as a variational set to compute the energy of excited states in a spectrum, a rule of thumb is to trust only circa the lowest half of the spectrum (in Ref. \cite{Bicudo:2016eeu}, among 11 to 12 states, only the lightest 6 are trusted). Just to illustrate why we must exclude the remaining states, in Fig. \ref{fig:pot_all} we show also the states we discard.   Notice the discarded states already have their spectrum saturated. 

Moreover these discarded states do not follow, at larger distances, the expected spectrum form the Nambu-Gotto string model \cite{Nambu:1978bd,Goto:1971ce}, expressed in the Arvis potential \cite{Luscher:1980iy,Arvis:1983fp},
\be
V_{n}(R) = \sigma\sqrt{R^{2}+\frac{2\pi}{\sigma}(n-\frac{D-2}{24})} \ ,
\label{Arvis}
\ee
where an infinite tower of excitations is predicted.
Our accepted five states correspond, at larger distances, to the Arvis potential with $n=0, \ 1, \ 2, \ 3$.

In what concerns comparing with the spectra of Ref. \cite{Juge:1999ie,Juge:2002br}, our states have similar energies. The only difference is in our highest states in the spectrum, our $\Pi^*_u$ is slightly heavier than our $\Sigma^+_u$, whereas Ref. \cite{Juge:1999ie,Juge:2002br} finds  $\Sigma^+_u$ slighly lighter than $\Pi^*_u$. Nevertheless, in both studies the differences between the spectrum of these two states are very small.

Thus, in the quantum number $\Sigma^+_g$ we accept two states, in the quantum number $\Pi_u$ we accept two states and in the quantum number $\Sigma^+_u$ we accept only one state. 

\begin{figure}[t!]
\begin{centering}
\includegraphics[width=\LagFigSize]{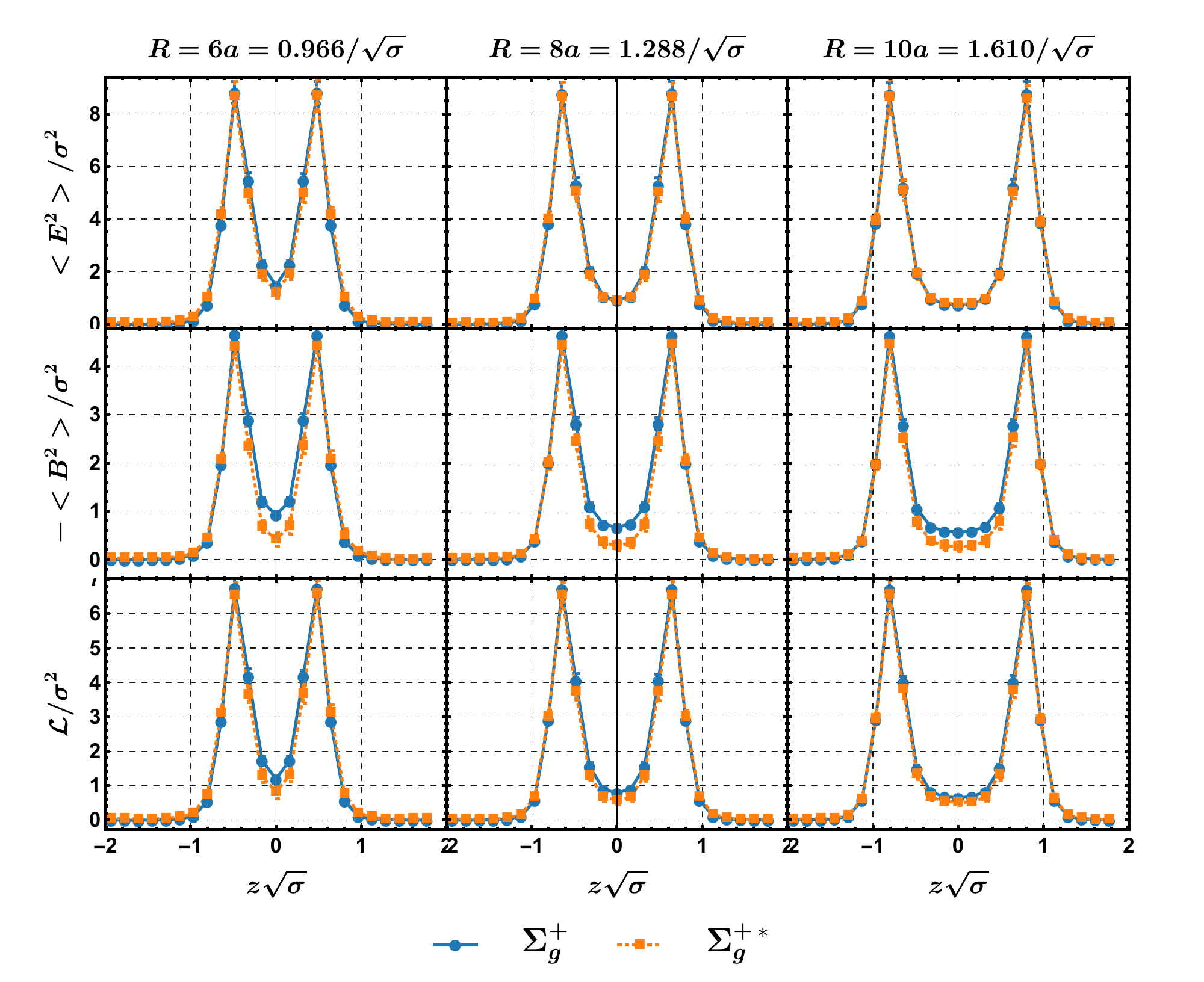}
\includegraphics[width=\LagFigSize]{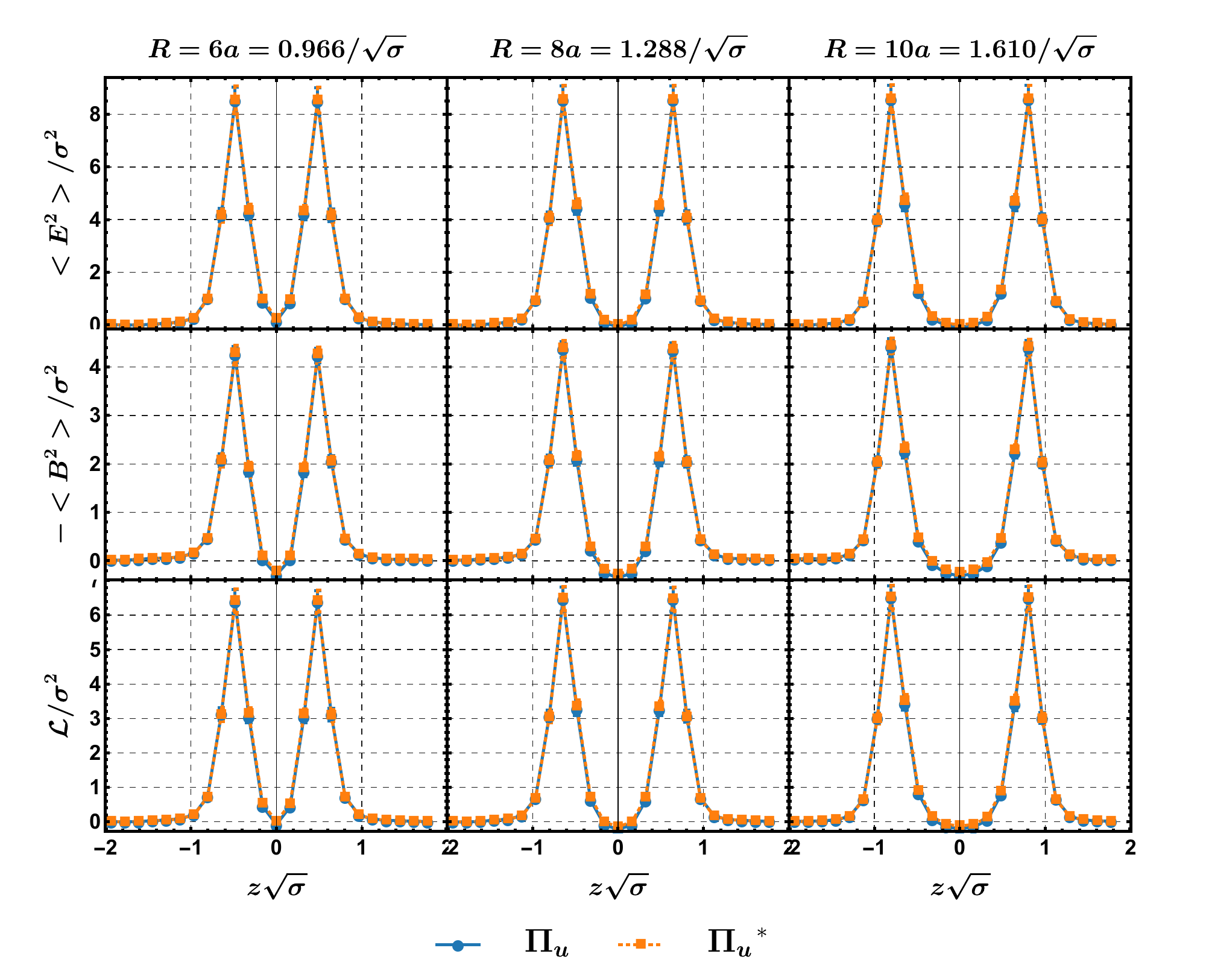}
\includegraphics[width=\LagFigSize]{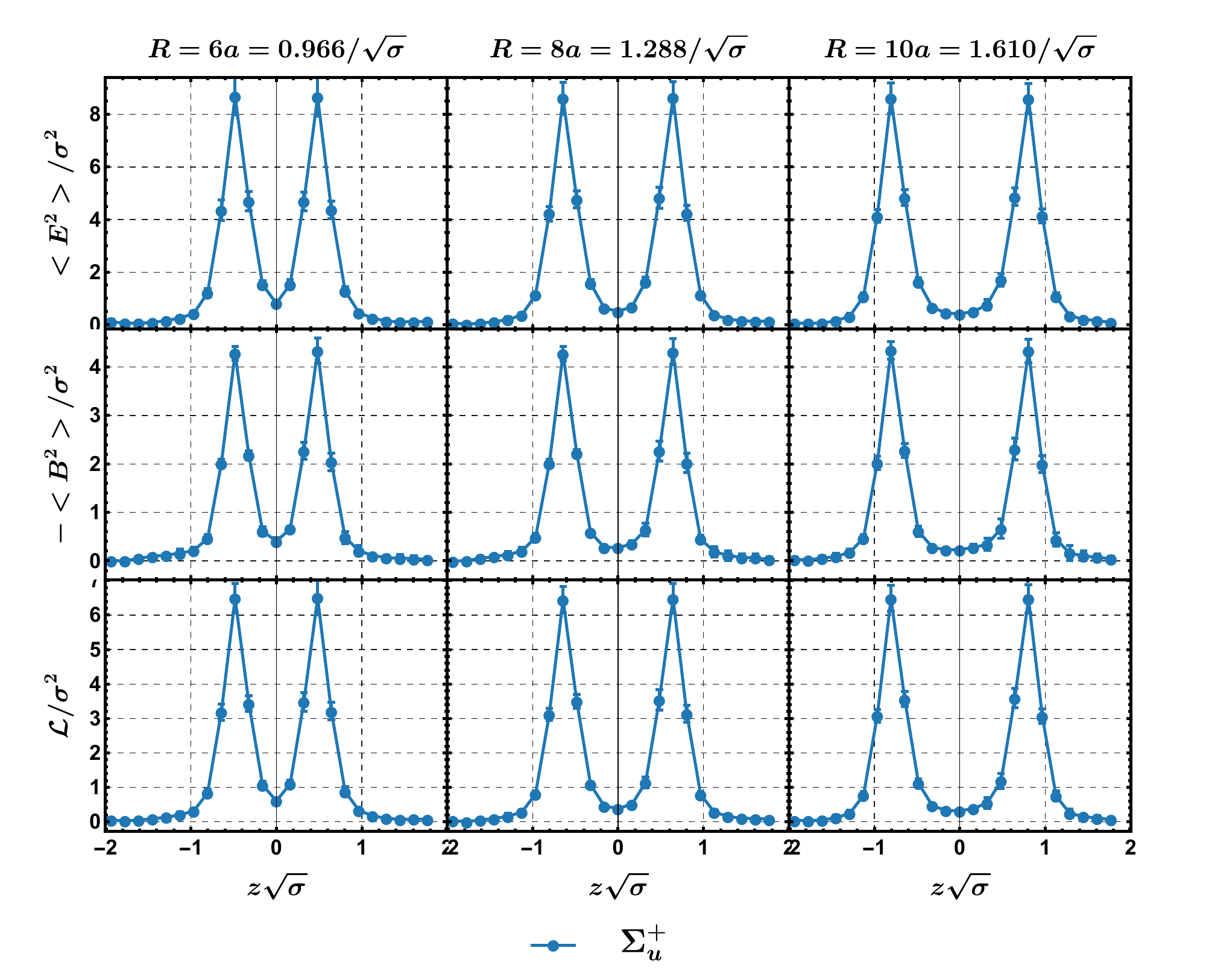}
\par\end{centering}
\caption{Lagrangian $\cal L$, $E^2$ and $B^2$ field densities in the charges axis. We show the groundstate and the excited states respectively for the quantum numbers $\Sigma^+_g$, $\Pi_u$ and $\Sigma^+_u$. The distance and energy are shown in string tension units $\sqrt \sigma$.
\label{fig:field_L_z}}
\end{figure}

\begin{figure}[t!]
\begin{centering}
\includegraphics[width=\LagFigSize]{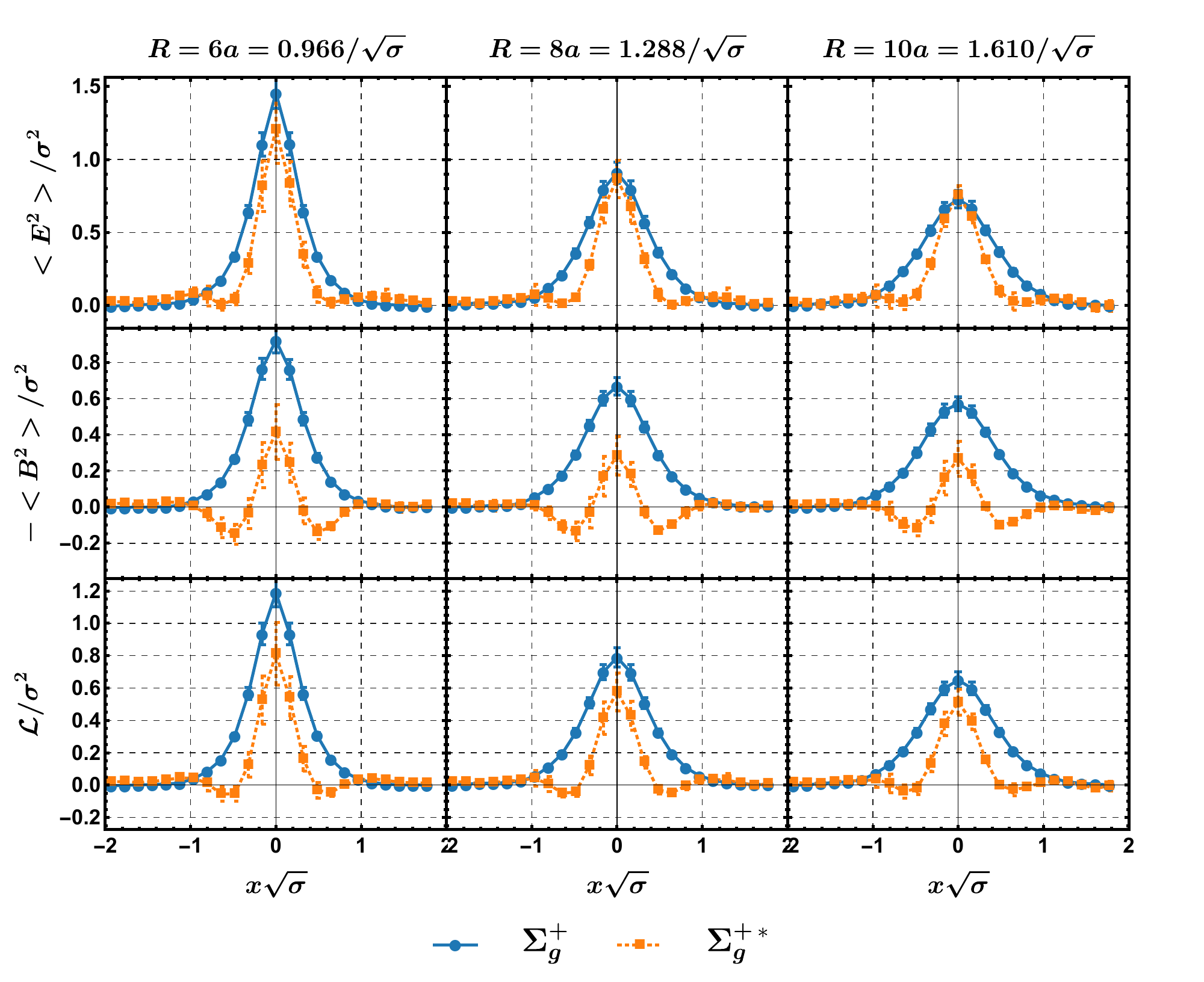}
\includegraphics[width=\LagFigSize]{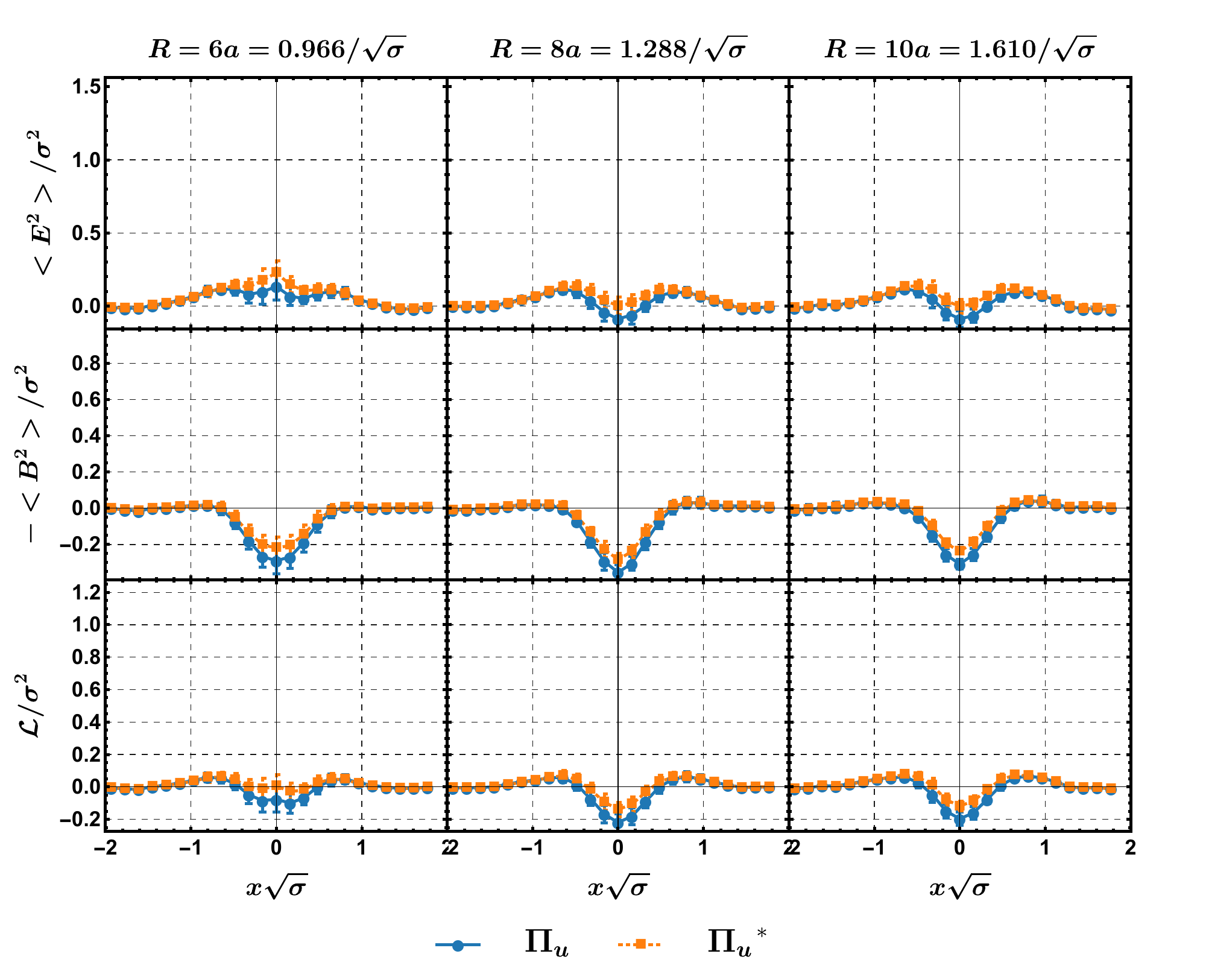}
\includegraphics[width=\LagFigSize]{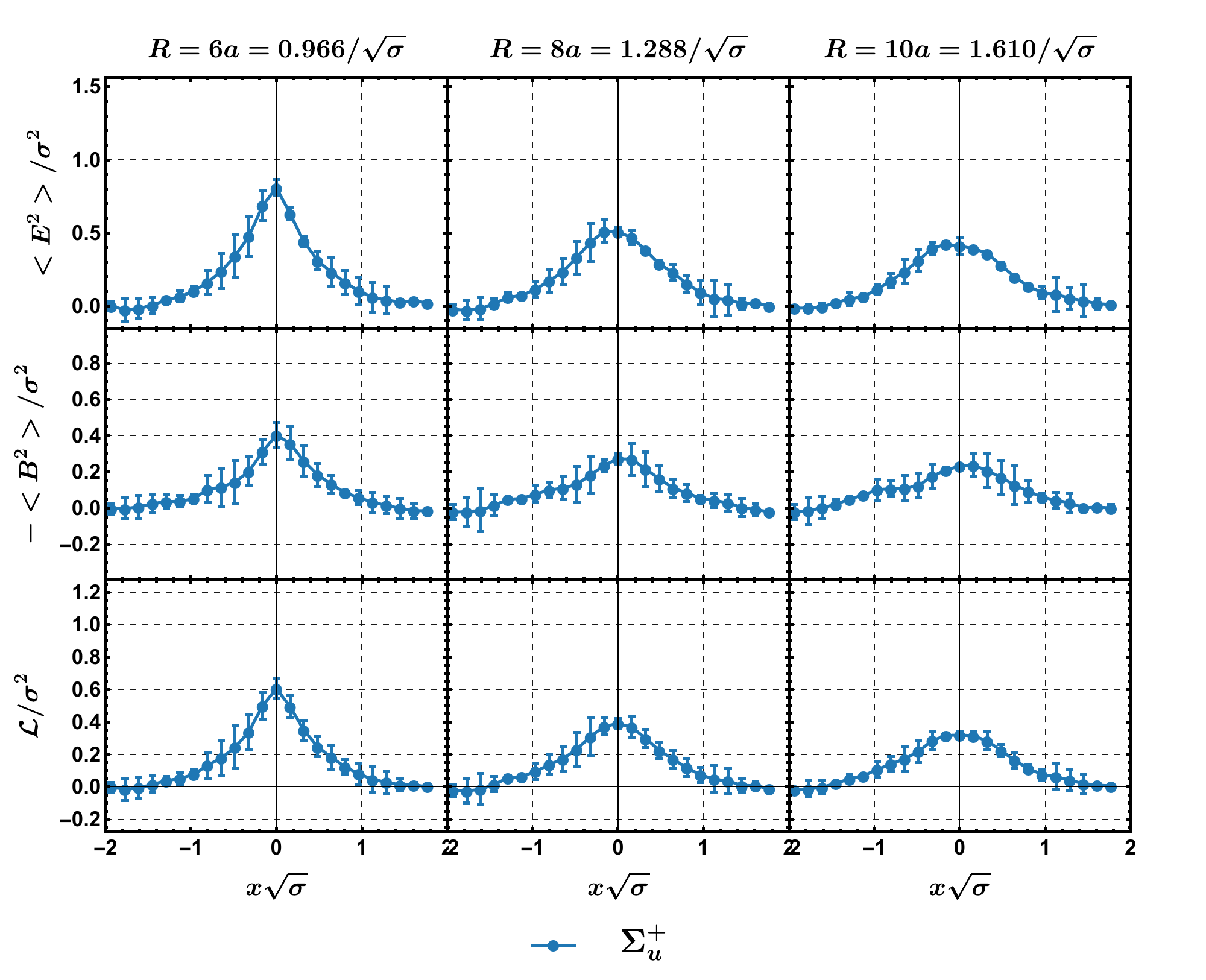}
\par\end{centering}
\caption{Lagrangian $\cal L$, $E^2$ and $B^2$ field densities in the mediator plane. We show the groundstate and the excited states respectively for the quantum numbers $\Sigma^+_g$, $\Pi_u$ and $\Sigma^+_u$. The distance and energy are shown in string tension units $\sqrt \sigma$.
\label{fig:field_L_xy}}
\end{figure}

\begin{figure}[t!]
\begin{centering}
\includegraphics[width=\LLagFigSize]{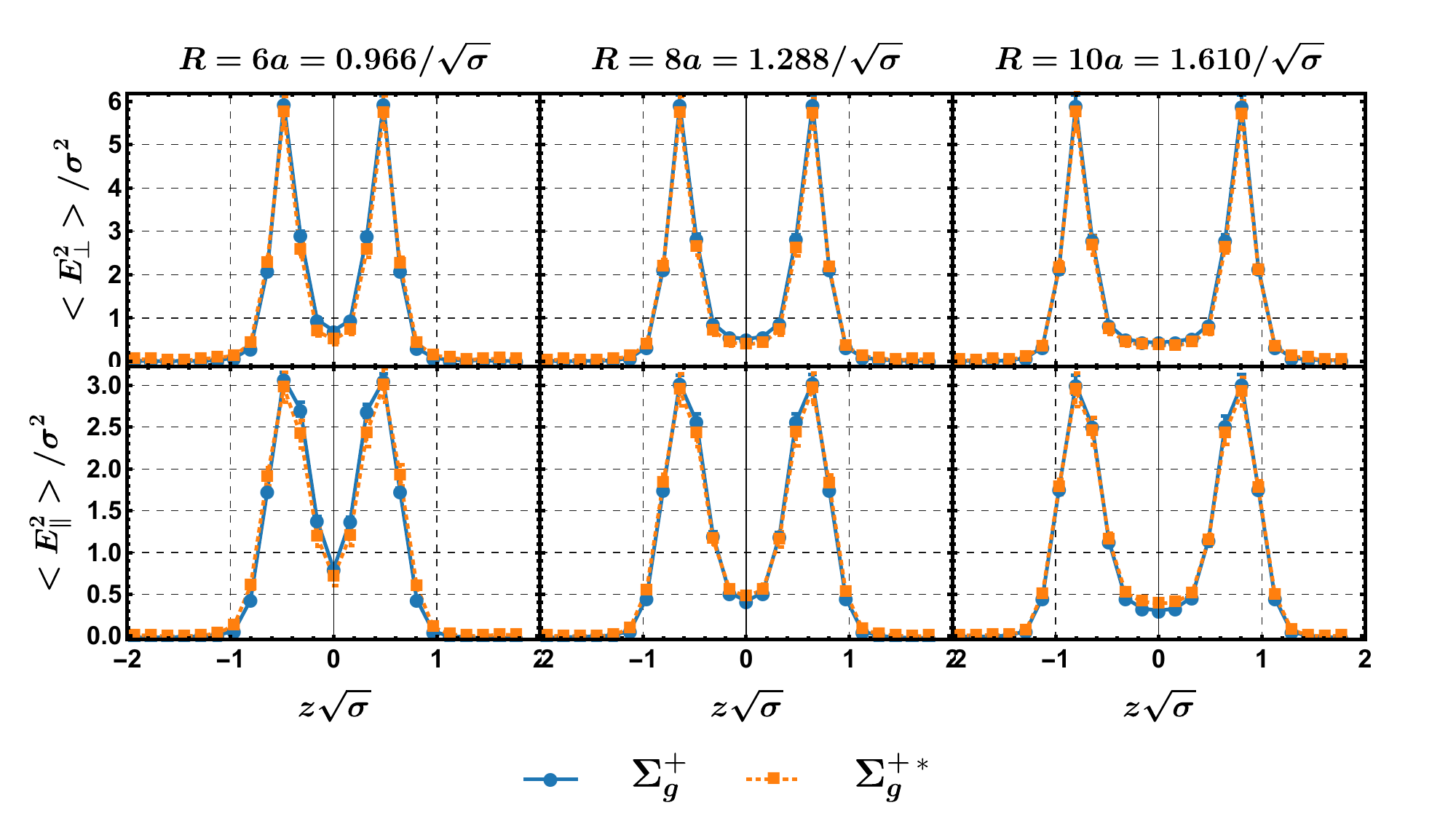}
\includegraphics[width=\LLagFigSize]{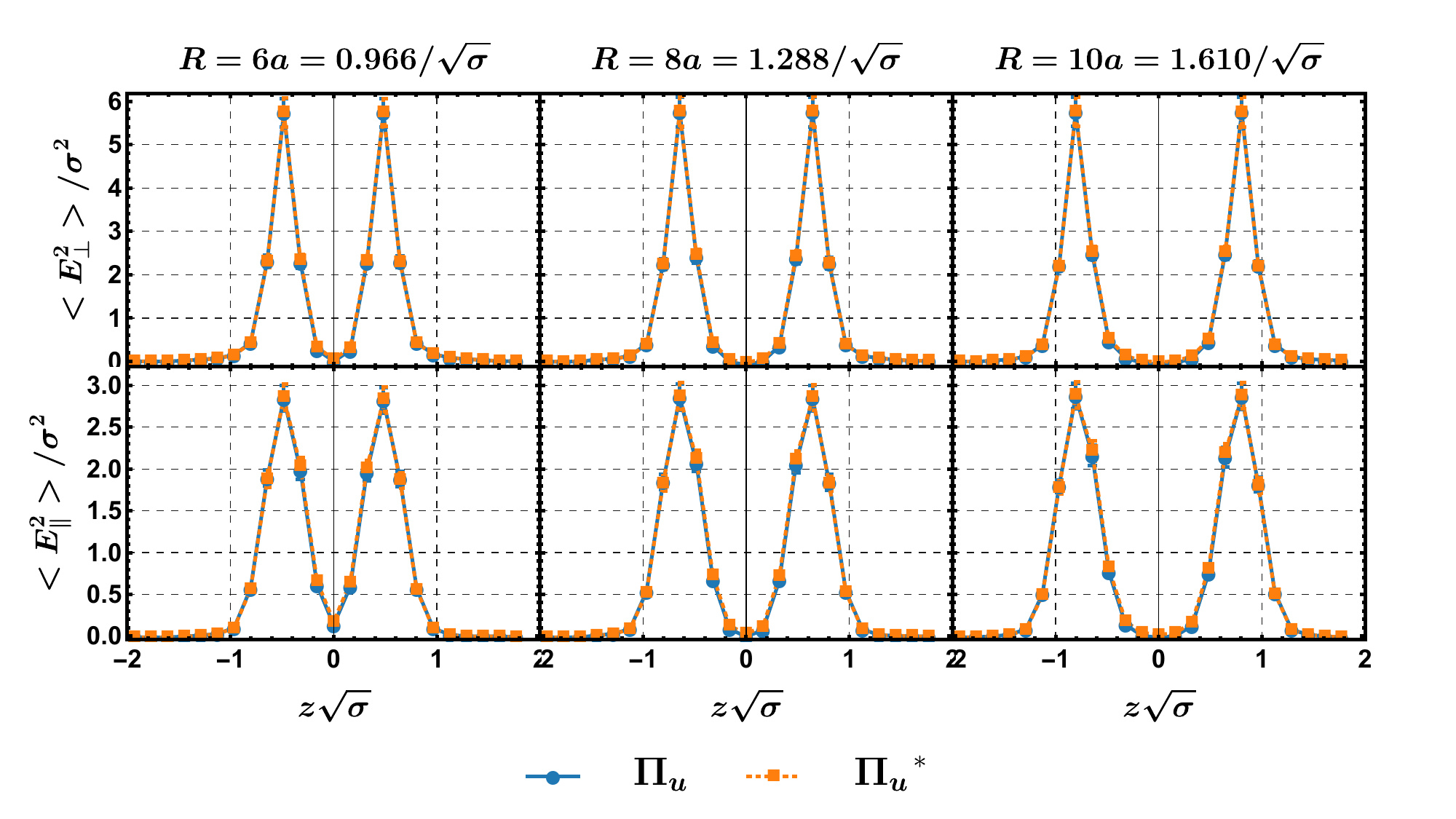}
\includegraphics[width=\LLagFigSize]{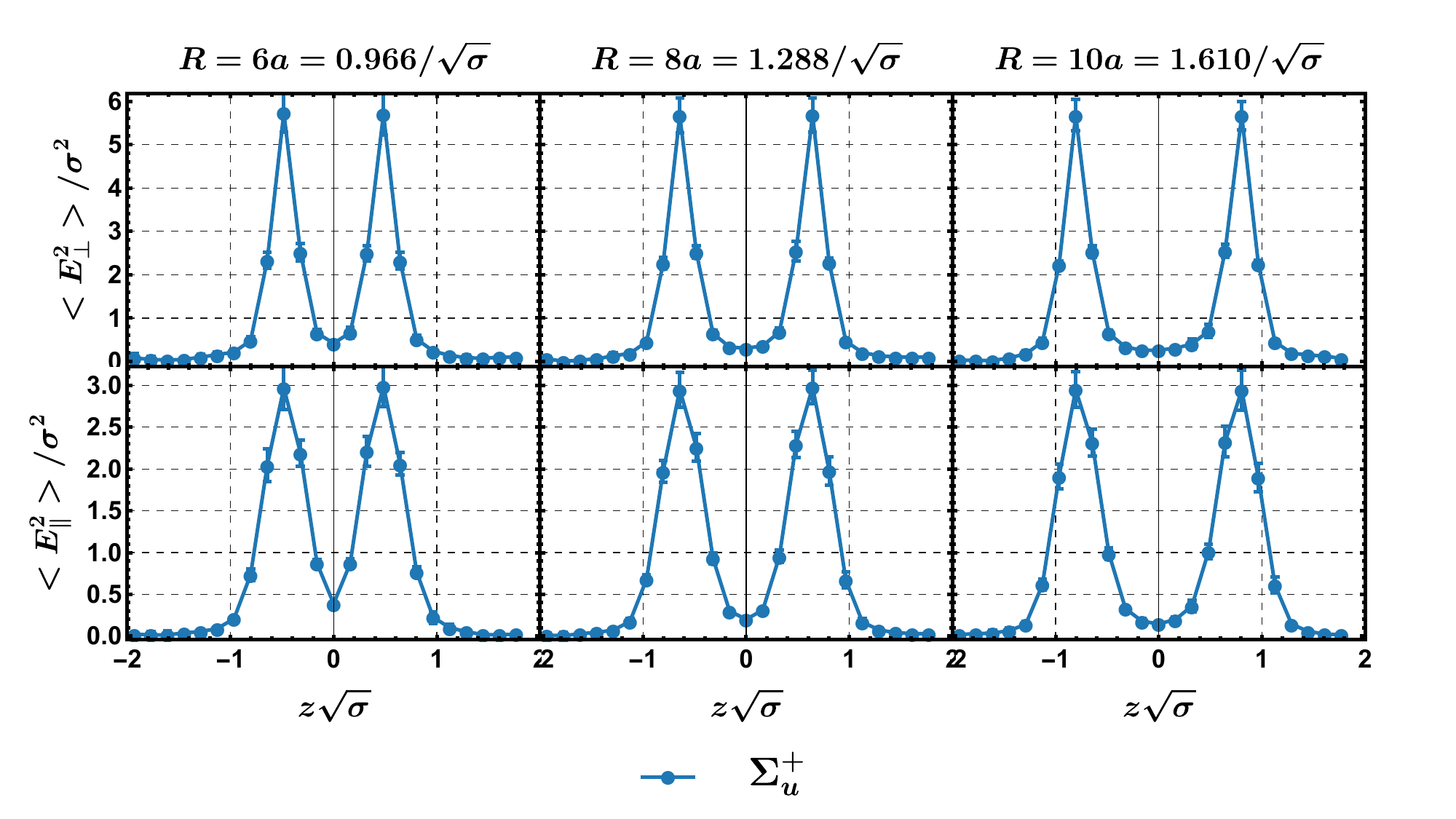}
\par\end{centering}
\caption{Chromoelectric ${E_\parallel}^2={E_z}^2$ and ${E_\perp}^2={E_x}^2+{E_y}^2$ component field densities in the charges axis. We show the groundstate and the excited states respectively for the quantum numbers $\Sigma^+_g$, $\Pi_u$ and $\Sigma^+_u$. The distance and energy are shown in string tension units $\sqrt \sigma$.
\label{fig:field_E_z}}
\end{figure}

\begin{figure}[t!]
\begin{centering}
\includegraphics[width=\LLagFigSize]{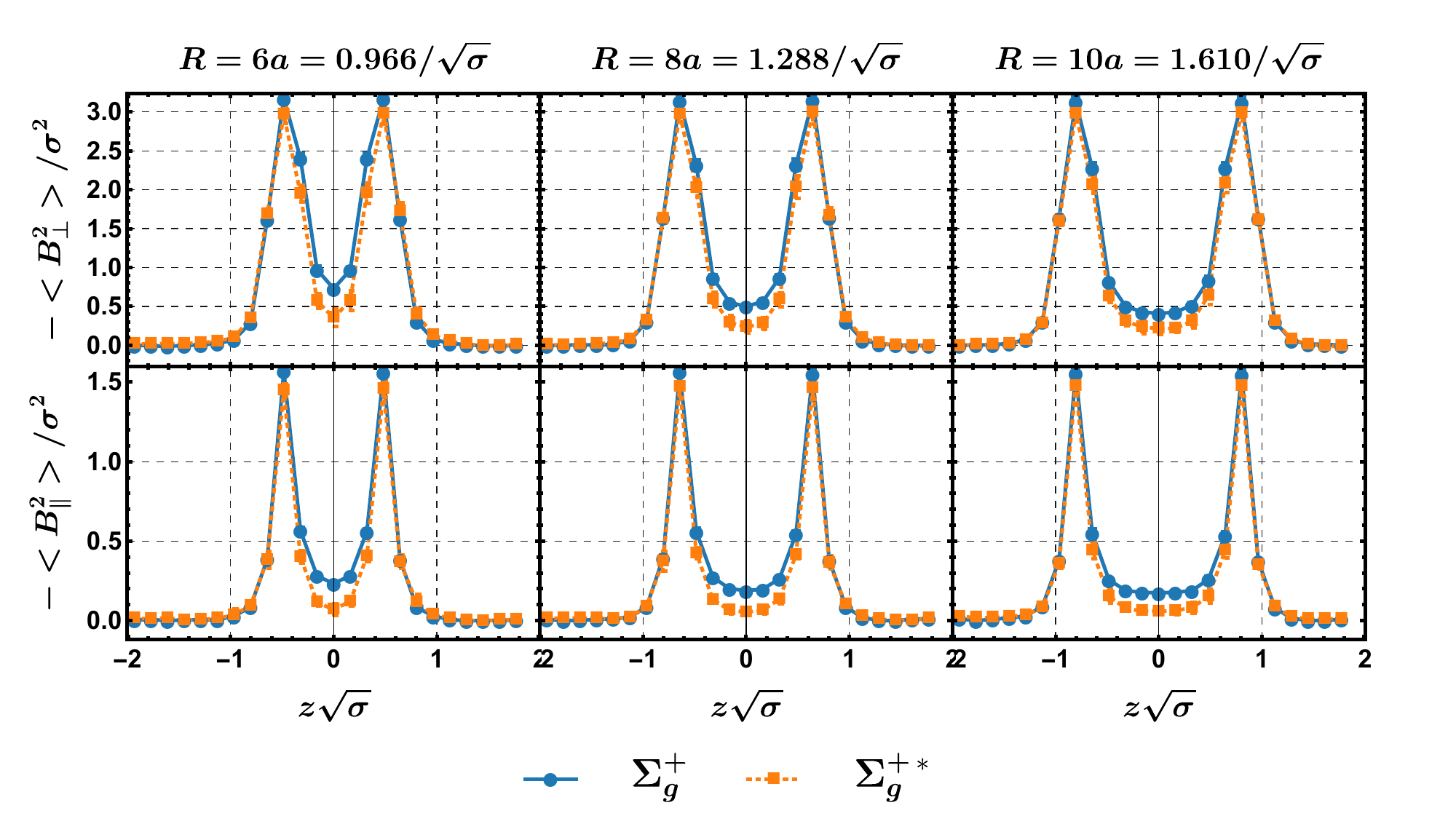}
\includegraphics[width=\LLagFigSize]{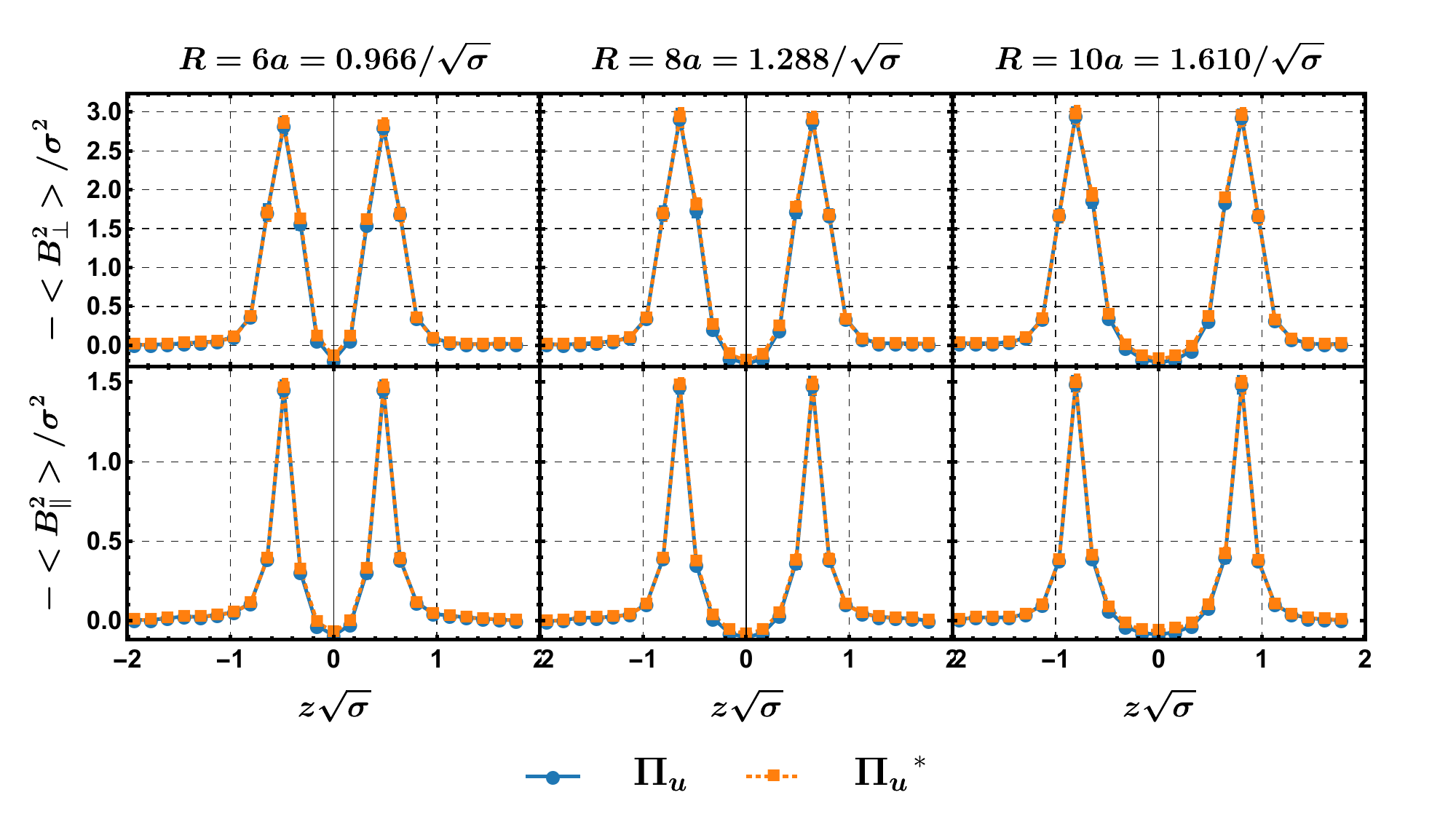}
\includegraphics[width=\LLagFigSize]{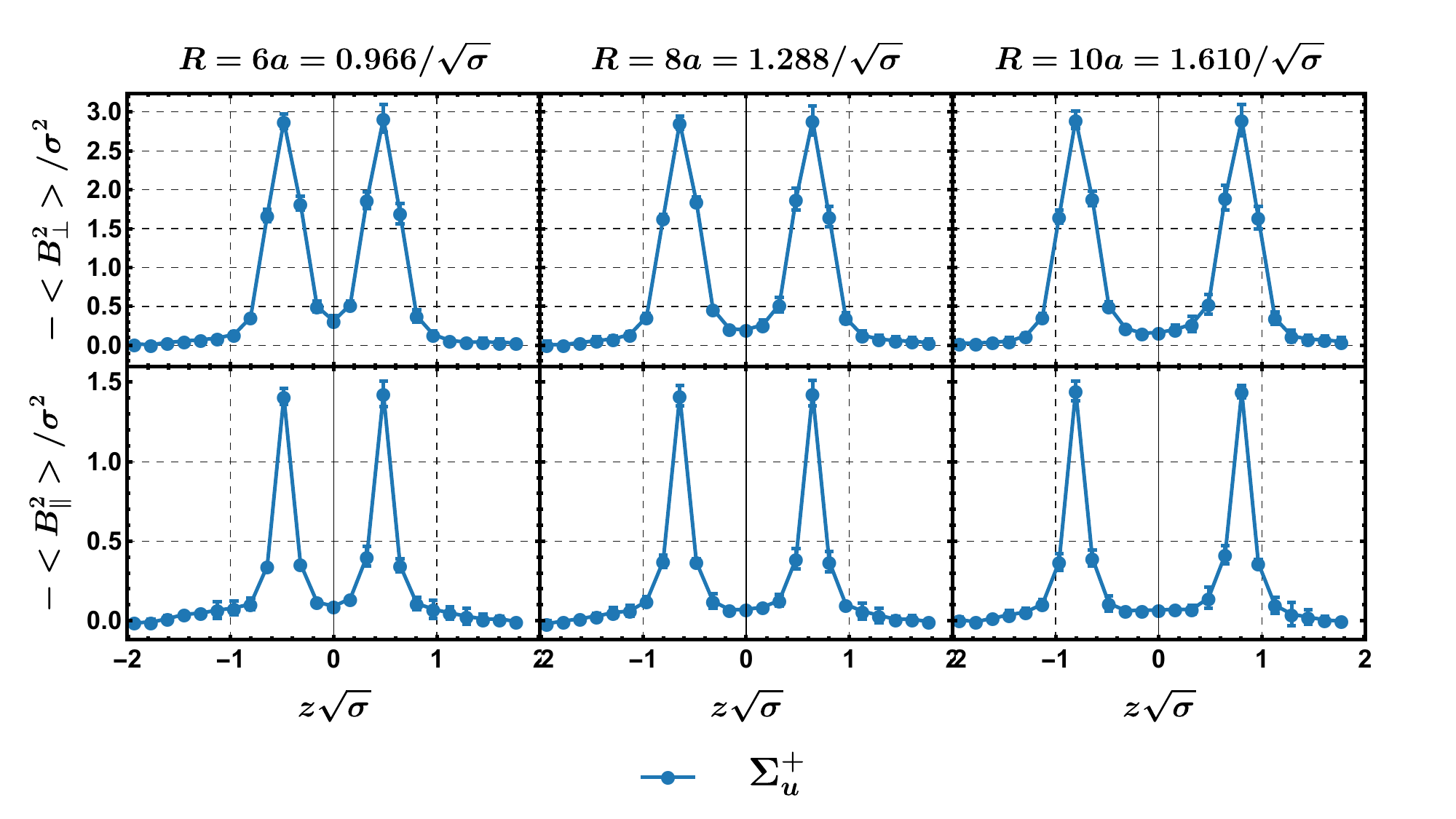}
\par\end{centering}
\caption{Chromomagnetic ${B_\parallel}^2$ and ${B_\perp}^2$ component field densities in the charges axis. We show the groundstate and the excited states respectively for the quantum numbers $\Sigma^+_g$, $\Pi_u$ and $\Sigma^+_u$. The distance and energy are shown in string tension units $\sqrt \sigma$.
\label{fig:field_B_z}}
\end{figure}

\begin{figure}[t!]
\begin{centering}
\includegraphics[width=\LLagFigSize]{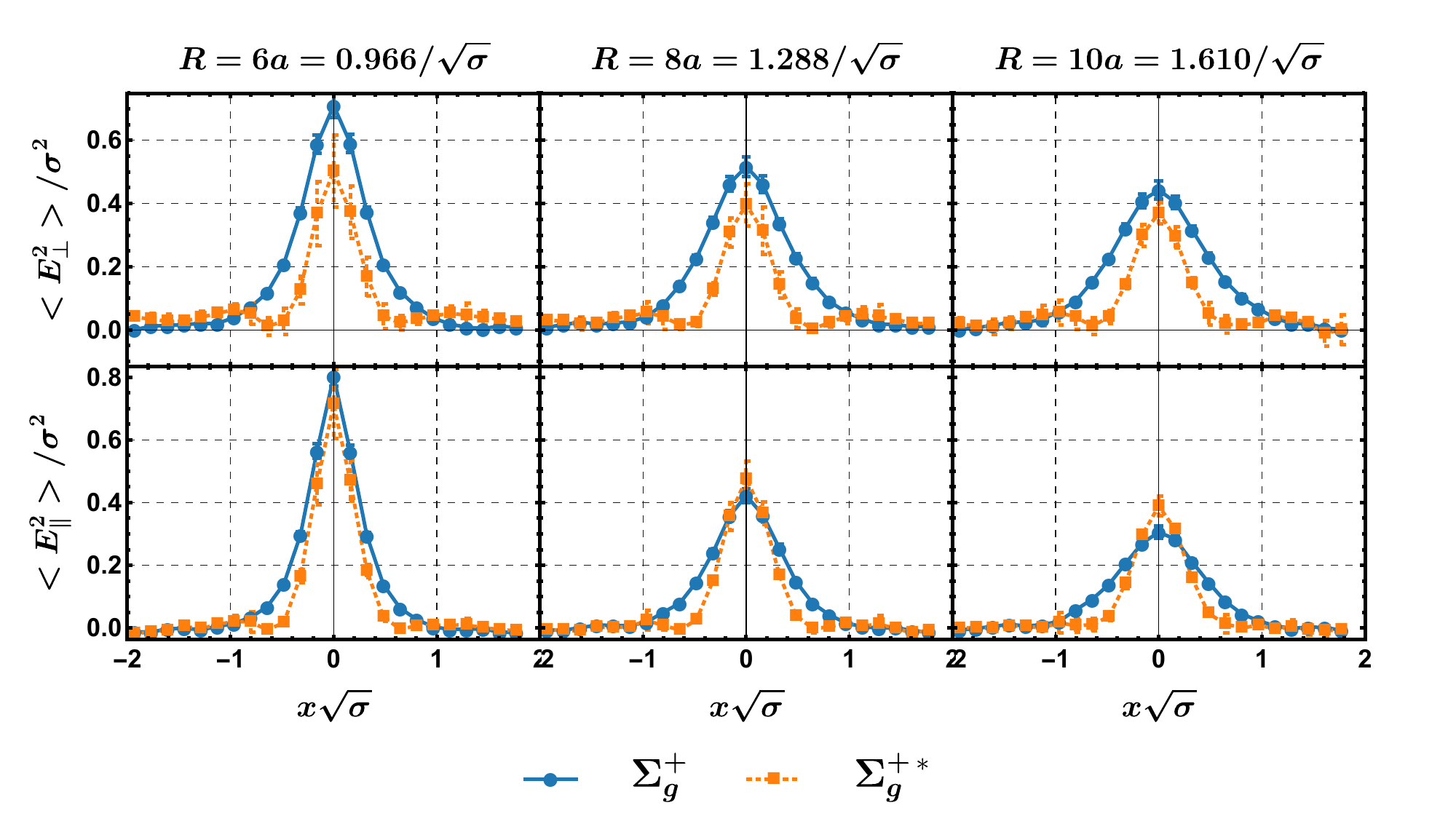}
\includegraphics[width=\LLagFigSize]{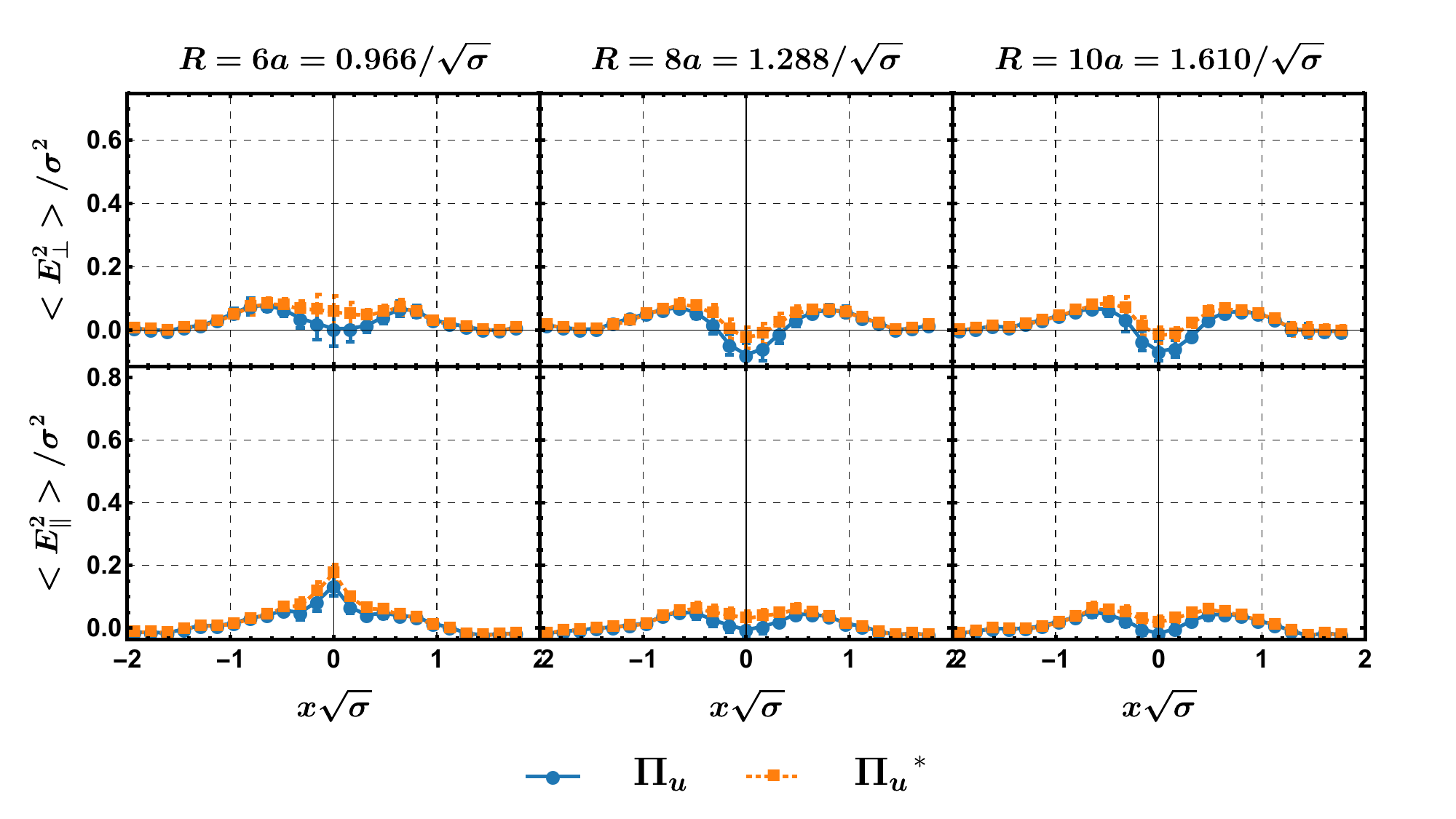}
\includegraphics[width=\LLagFigSize]{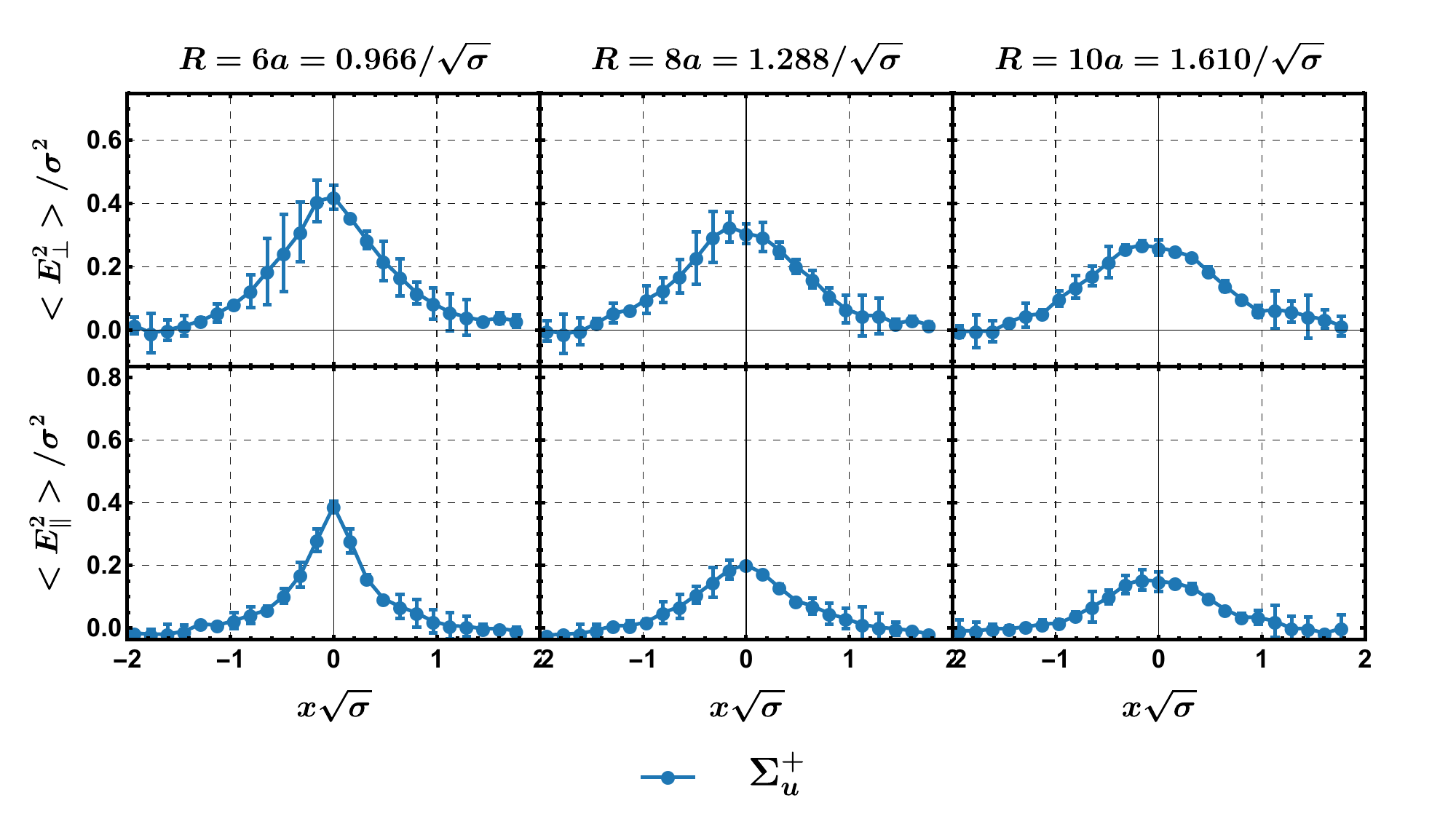}
\par\end{centering}
\caption{Chromoelectric ${E_\parallel}^2$ and ${E_\perp}^2$ component field densities in the mediator plane. We show the groundstate and the excited states respectively for the quantum numbers $\Sigma^+_g$, $\Pi_u$ and $\Sigma^+_u$. The distance and energy are shown in string tension units $\sqrt \sigma$.
\label{fig:field_E_xy}}
\end{figure}

\begin{figure}[t!]
\begin{centering}
\includegraphics[width=\LLagFigSize]{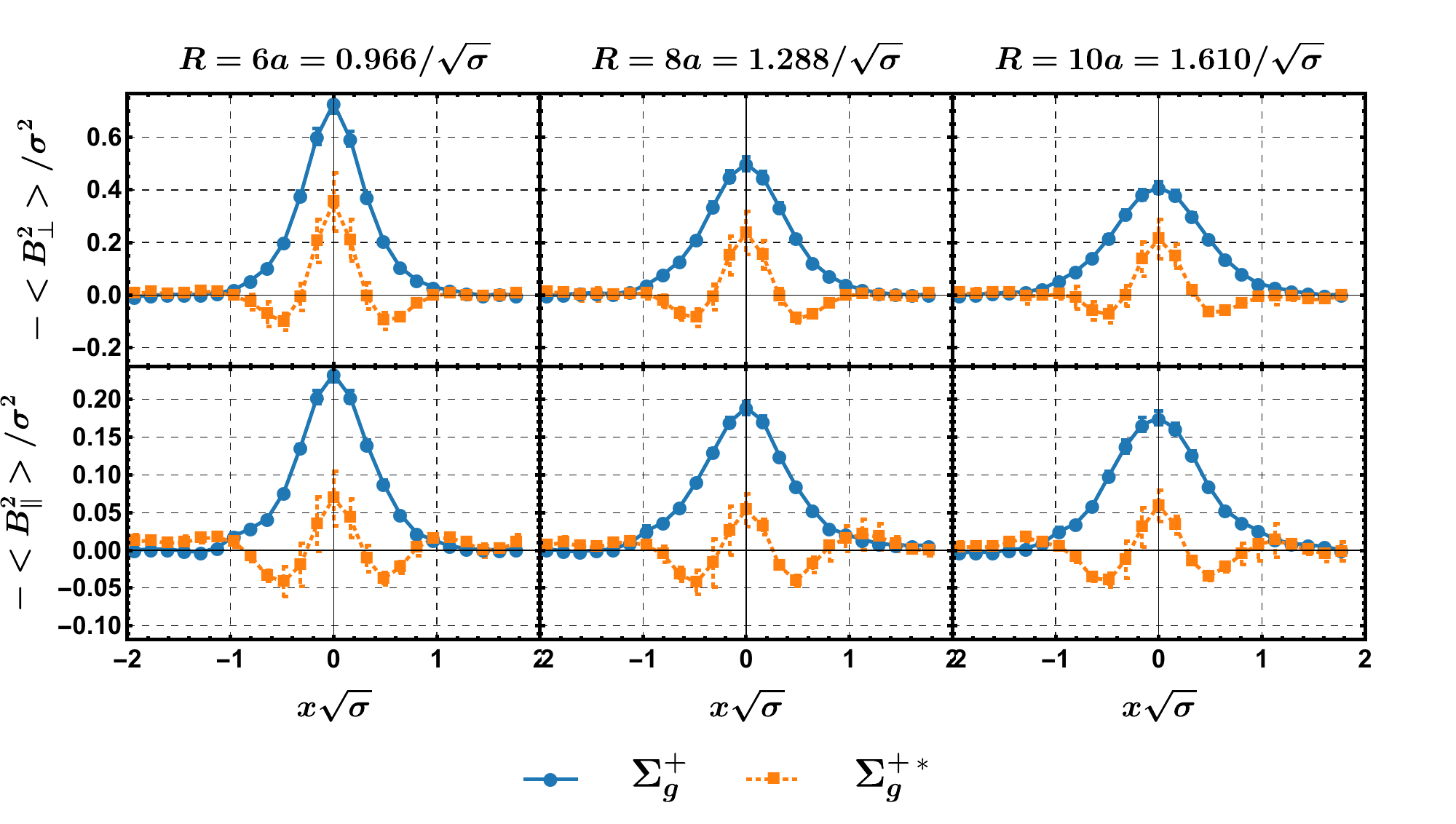}
\includegraphics[width=\LLagFigSize]{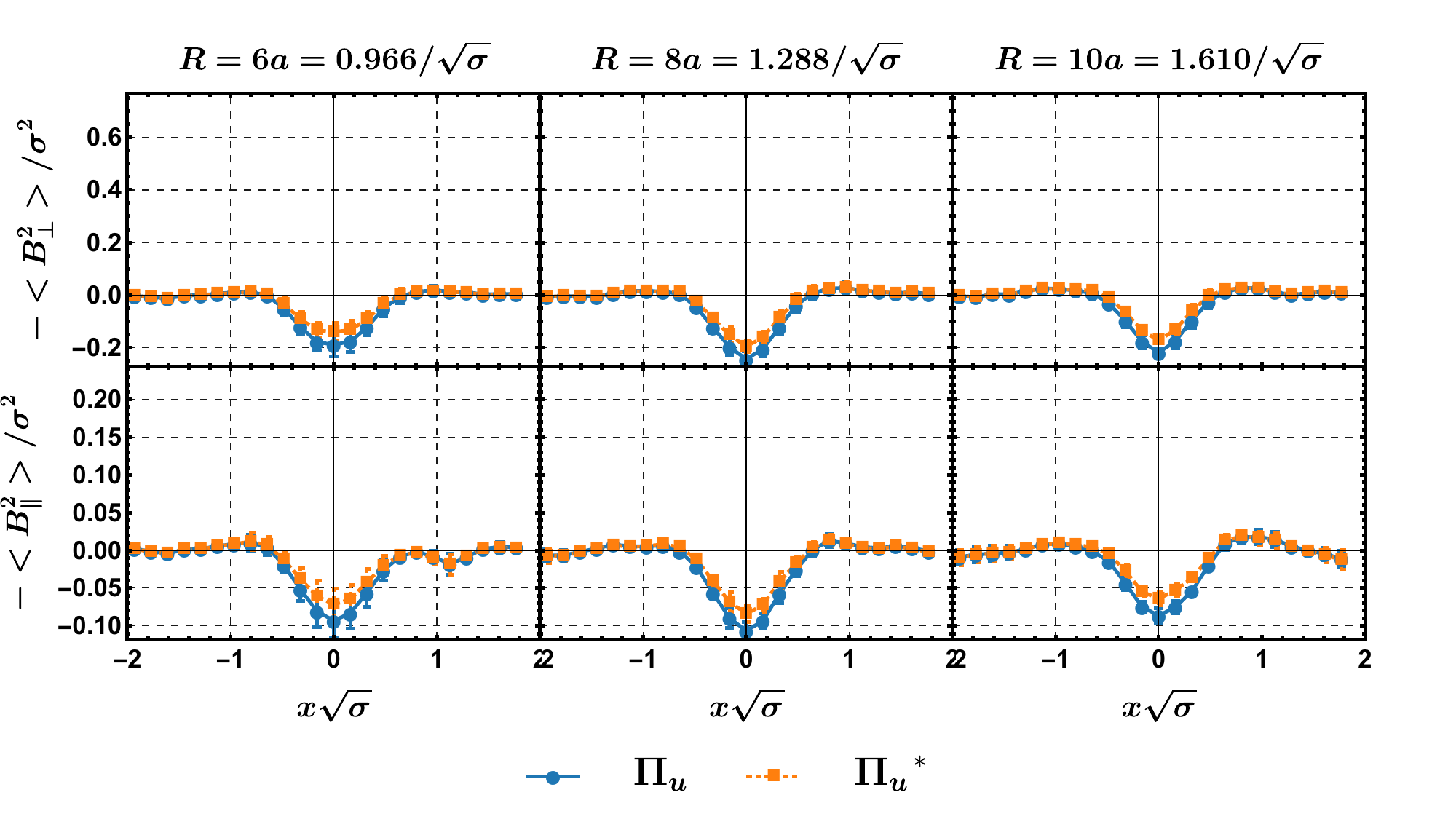}
\includegraphics[width=\LLagFigSize]{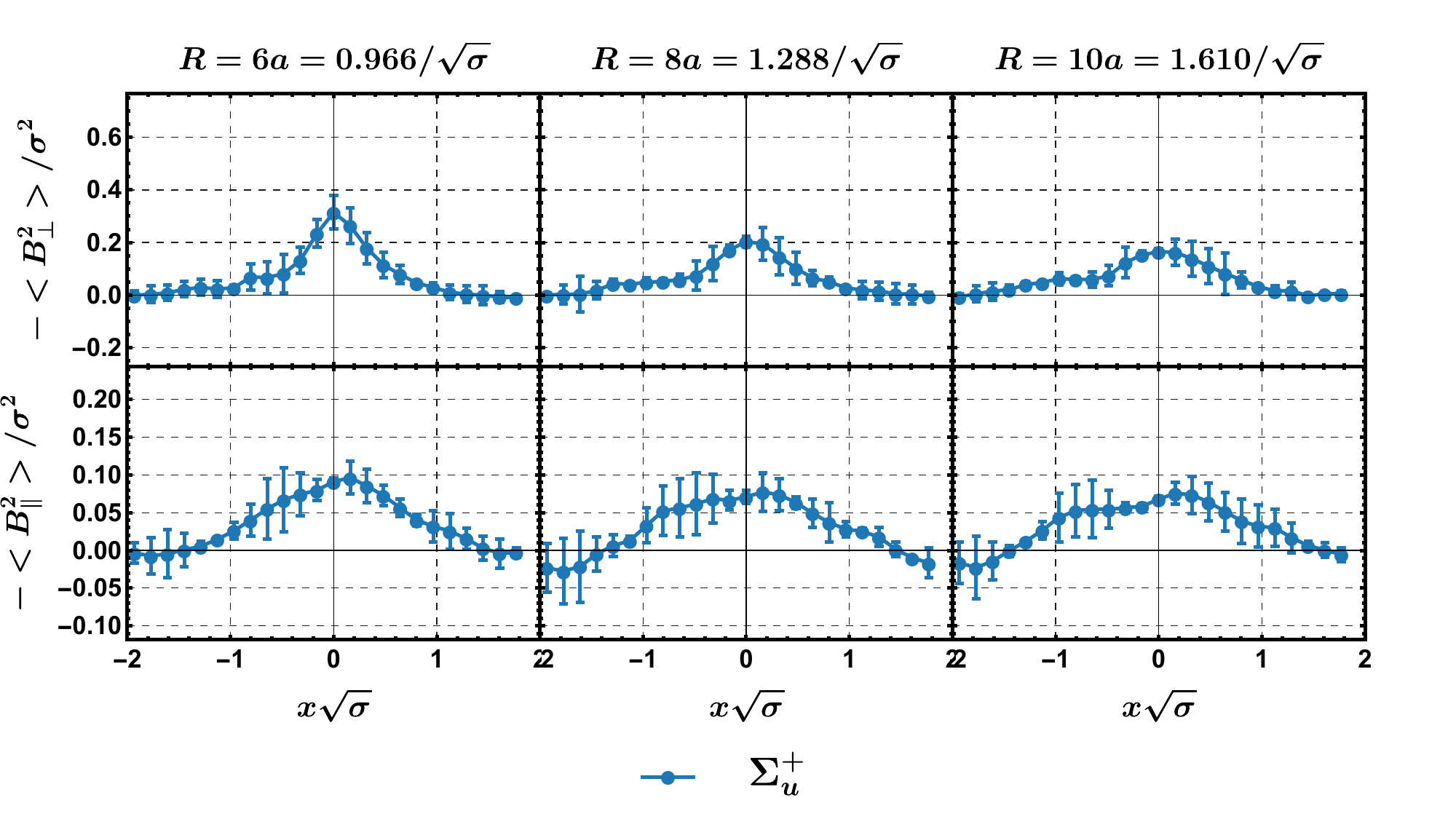}
\par\end{centering}
\caption{Chromomagnetic ${B_\parallel}^2$ and ${B_\perp}^2$ component field densities in the mediator plane. We show the groundstate and the excited states respectively for the quantum numbers $\Sigma^+_g$, $\Pi_u$ and $\Sigma^+_u$. The distance and energy are shown in string tension units $\sqrt \sigma$.
\label{fig:field_B_xy}}
\end{figure}

Our results for the flux tubes are presented in Figs.  \ref{fig:field_L_z}, \ref{fig:field_L_xy}, \ref{fig:field_E_z}, \ref{fig:field_B_z}, \ref{fig:field_E_xy} and \ref{fig:field_B_xy}. In all these figures we show the flux tubes for the groundstate  $\Sigma^+_g$ and its first excitation (s-wave, parity + for the charge conjugation and inversion) and its excitations, the flux tubes for the $\Pi_u$ and its first excitation (p-wave,  parity - for the charge conjugation and inversion) and the  flux tubes for the $\Sigma^+_u$ (s-wave, parity -   for the charge conjugation and inversion).

In particular we show in  Fig.  \ref{fig:field_L_z} and in Fig.  \ref{fig:field_L_xy} the Lagrangian field density $\cal L$, the electric field density $E^2$ and  the magnetic field density $B^2$ respectively, both in the charges axis and in the mediator plane.

In Fig.  \ref{fig:field_E_z} we show the components of chromoelectric field density in the charges axis,  in Fig.  \ref{fig:field_B_z}  the components of the chromomagnetic field density in the charges axis,  in Fig. \ref{fig:field_E_xy}  the components of the chromoelectric field density in the mediator plane, and  in Fig. \ref{fig:field_B_xy}  the components of the chromomagnetic field density in the mediator plane.  We separate the parallel ${E_\parallel}^2={E_z}^2$, ${B_\parallel}^2={B_z}^2$ and the perpendicular ${E_\perp}^2={E_x}^2+{E_y}^2$, ${B_\perp}^2={B_x}^2+{B_y}^2$ components.

In Figs. \ref{fig:3dplot_L_xy} and \ref{fig:3dplot_PIu_xy} we analyse in 3D plots the density profile in the whole mediator plane.

\section{Analysis of the flux tubes \label{sec:analysis}}

In this first exploratory study of flux tubes, we analyse the difference between quantum numbers $\Sigma \ / \ \Pi$ and  $g \ / \ u$. Notice how the first excited state ${\Sigma^+_g}^*$ differs from the groundstate  $\Sigma^+_g$. Its profile in the mediator plane has an extra node, as expected in a radial excitation.

We  now search for evidences of phenomena beyond the bosonic Nambu-Goto string model.

%

%

%

%

%

%

\subsection{Fields in the charges neighbourhood}

The clearest difference between the quantum string model and the flux tubes is in the case of short inter-charge distance $R$. Not only the groundstate potential has no tachyon, unlike the Jarvis potential, but also the fields of the charges are very large. This is consistent with the onset of perturbative-like QCD at short distances 
\cite{Karbstein:2014bsa}
and with Coulomb potentials.

\subsection{Densities $E^2_\perp$, $E^2_\parallel$, $B^2_\perp$, $B^2_\parallel$ \label{sec:components}}

Unlike the bosonic string model, the fields in the flux tubes have several components. 
We show the different components,  the parallel ${E_\parallel}^2={E_z}^2$, ${B_\parallel}^2={B_z}^2$ and the perpendicular ${E_\perp}^2={E_x}^2+{E_y}^2$, ${B_\perp}^2={B_x}^2+{B_y}^2$ components in Figs. \ref{fig:field_E_z}, \ref{fig:field_B_z}, \ref{fig:field_E_xy} and \ref{fig:field_B_xy}. 

In Figs. \ref{fig:field_E_z} and \ref{fig:field_B_z} it is clear that, in the neighbourhood of the charges, each one of the squared Cartesian components of the chromoelectric fields $E^2_x, \ E^2_y, \ E^2_z$ has the same magnitude. Besides, the squared Cartesian components of the chromomagnetic  fields $B^2_x, \ B^2_y, \ B^2_z$ have a value of approximately $1/2$ of the chromoelectric ones. 

To study in more detail the flux tube, we also analyse the mediator plane in Figs.  \ref{fig:field_E_xy} and \ref{fig:field_B_xy}. There, the parallel chromoelectric field density ${E_\parallel}^2$ is of the order of the perpendicular one ${E_\perp}^2$, although the perpendicular component has two Cartesian coordinates. 

This goes approximately in the direction of the dual superconductor picture, where it would be expected that the ${E_\parallel}^2$ would absolutely dominate.

However the dual superconductor picture is not exact, since all different components are non-vanishing, including a large chromomagnetic  ${B_\perp}^2$.

\subsection{Searching for transverse versus longitudinal degrees of freedom.}

An effect beyond the string model would be the proof of longitudinal vibration modes. It is subtle to detect these modes.

We decided, for a first study with long non-local modes, to use in this work the operators of Fig. \ref{fig:basisex}, with a length of $R/2$. In this framework, we designed the circular operators $W^L_x$, $W^R_x$, $W^L_y \cdots$ to enhance the signal of the longitudinal degrees of freedom, exciting the components $B_\perp$, correlated in principle with the longitudinal $E_\parallel$ (the largest components as discussed in subsection \ref{sec:components}).
However these operators did not produce any observable improvement in any of the flux tubes we measured.

\begin{figure}[t!]
\begin{centering}
\includegraphics[width=\LagFigSize]{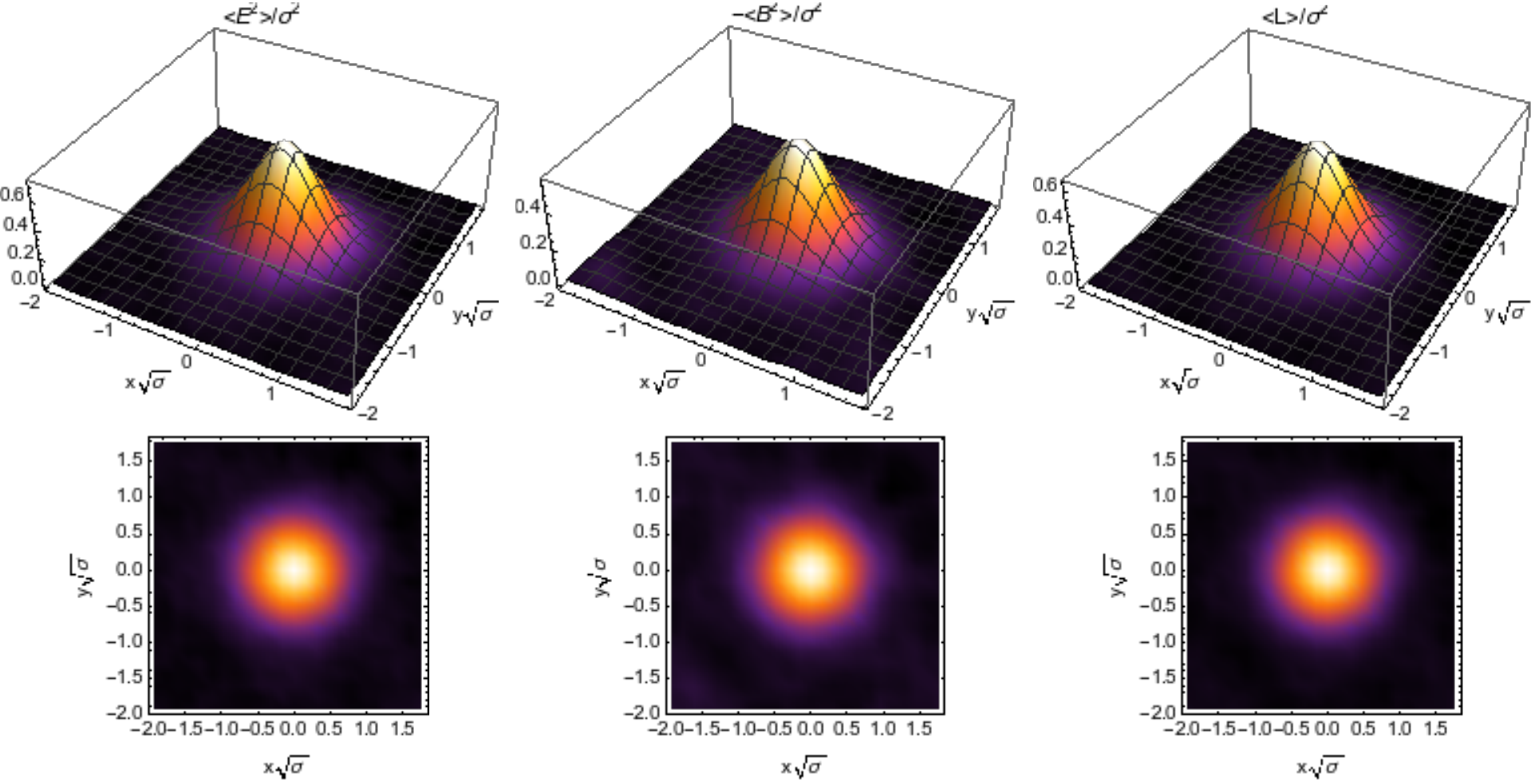}
\includegraphics[width=\LagFigSize]{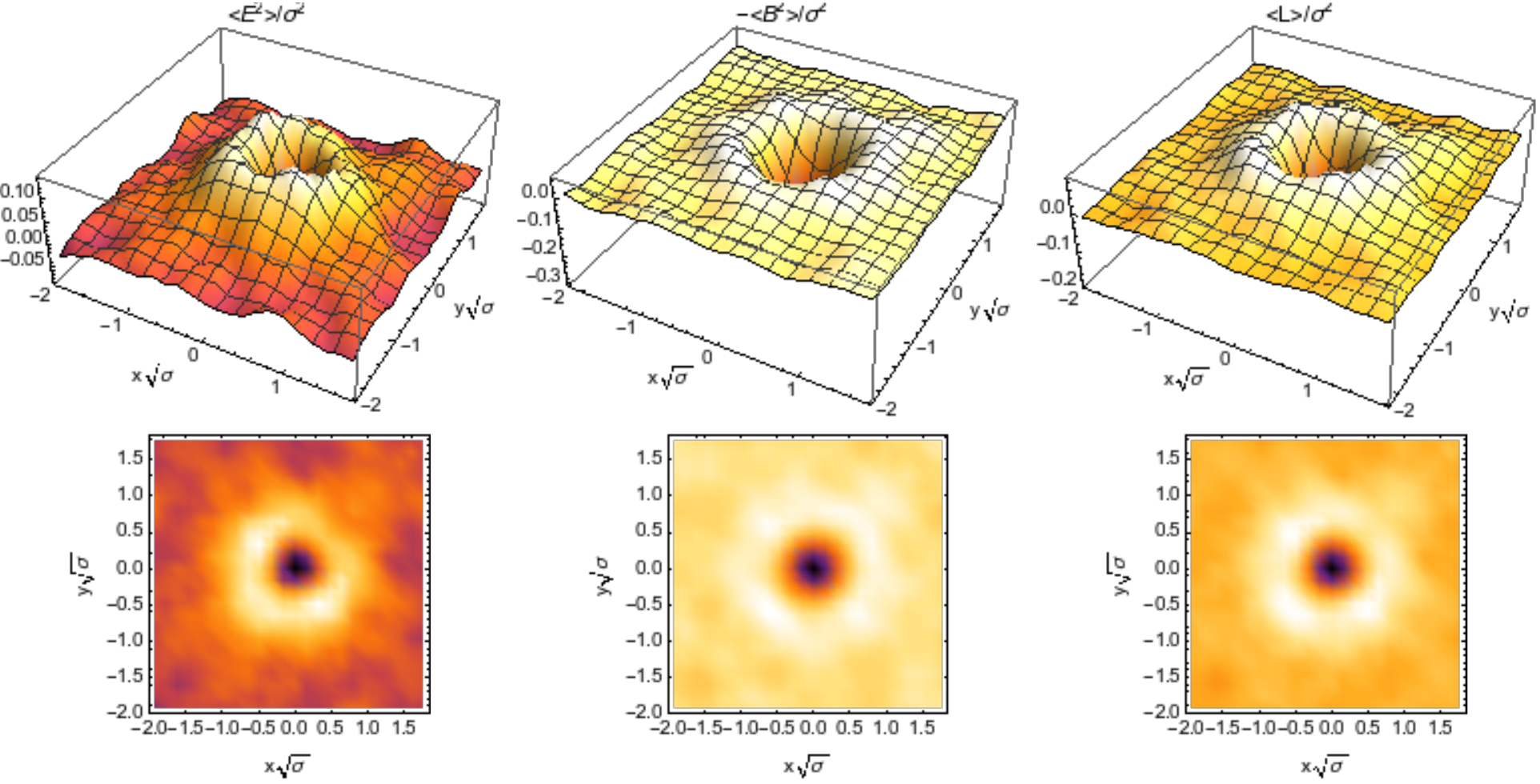}
\includegraphics[width=\LagFigSize]{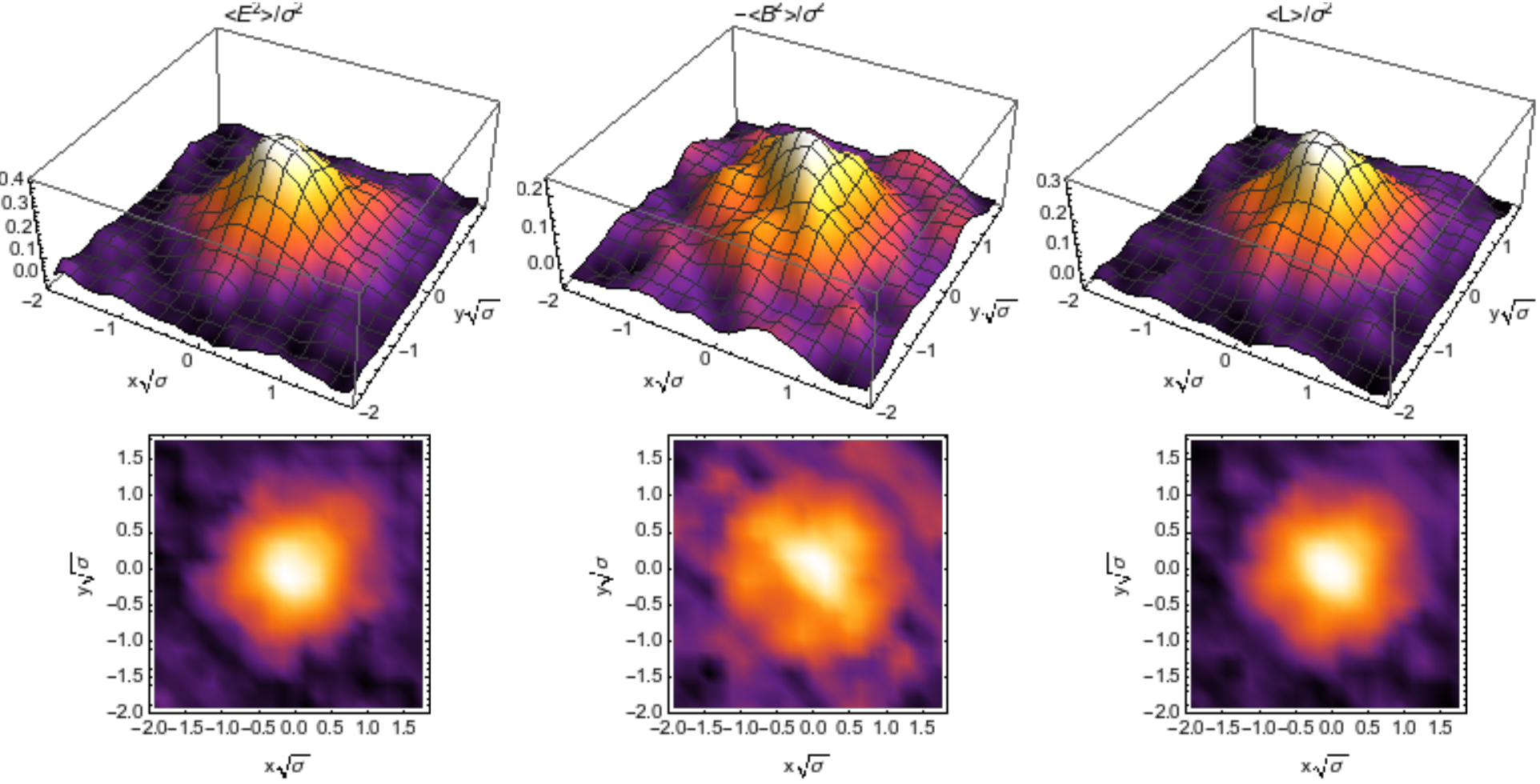}
\par\end{centering}
\caption{3D plots of the Lagrangian $\cal L$, $E^2$ and $B^2$ field densities in the mediator plane, for an inter-charge distance $R=10 a= 1.610 / \sqrt \sigma $. We show the groundstate respectively for the quantum numbers $\Sigma^+_g$ (top), $\Pi_u$ (centre) and $\Sigma^+_u$ (bottom). The distance and energy are shown in string tension units $\sqrt \sigma$. The magnitude of the densities is the same of Fig. \ref{fig:field_L_xy}.
\label{fig:3dplot_L_xy}}
\end{figure}

Another possible evidence of longitudinal waves would be in longitudinal quantum fluctuations. There we have some evidence in the $\Sigma^+_u$ fluxtubes. These flux tubes have parity $-$ and thus the field components should vanish in the median point of the flux tube, at $z=0$, as illustrated in Fig. \ref{fig:quantumnumbers}. However the squared Cartesian components $E^2_i$ and $B^2_i$ do not vanish in the mediator plane, and this may be interpreted as an evidence for longitudinal fluctuations of the flux tube. This is clear in Fig. \ref{fig:3dplot_L_xy}, where the third set of 3D plots, for the $\Sigma^+_u$ fluxtubes do not vanish in the mediator plane.

As a final evidence, in Figs.  \ref{fig:field_E_xy} and \ref{fig:field_B_xy}, the density for the parallel components ${E_\parallel}^2$, ${B_\parallel}^2$ do not vanish.
Notice we measure not only the fields but also their fluctuations.  If the flux tube would correspond to a transverse standing wave, the parallel components ${E_\parallel}^2$, ${B_\parallel}^2$ should vanish as they correspond to longitudinal fluctuations.

%

\subsection{Searching for the evidence of an explicit gluon}

An explicit gluon \cite{Mueller:2018fkg} would clearly go beyond the bosonic string model. Such a particle, a vector with parity $-$ is expected to be visible in the flux tube with quantum numbers $\Pi_u$.  

\begin{figure}[t!]
\begin{centering}
\includegraphics[width=\LagFigSize]{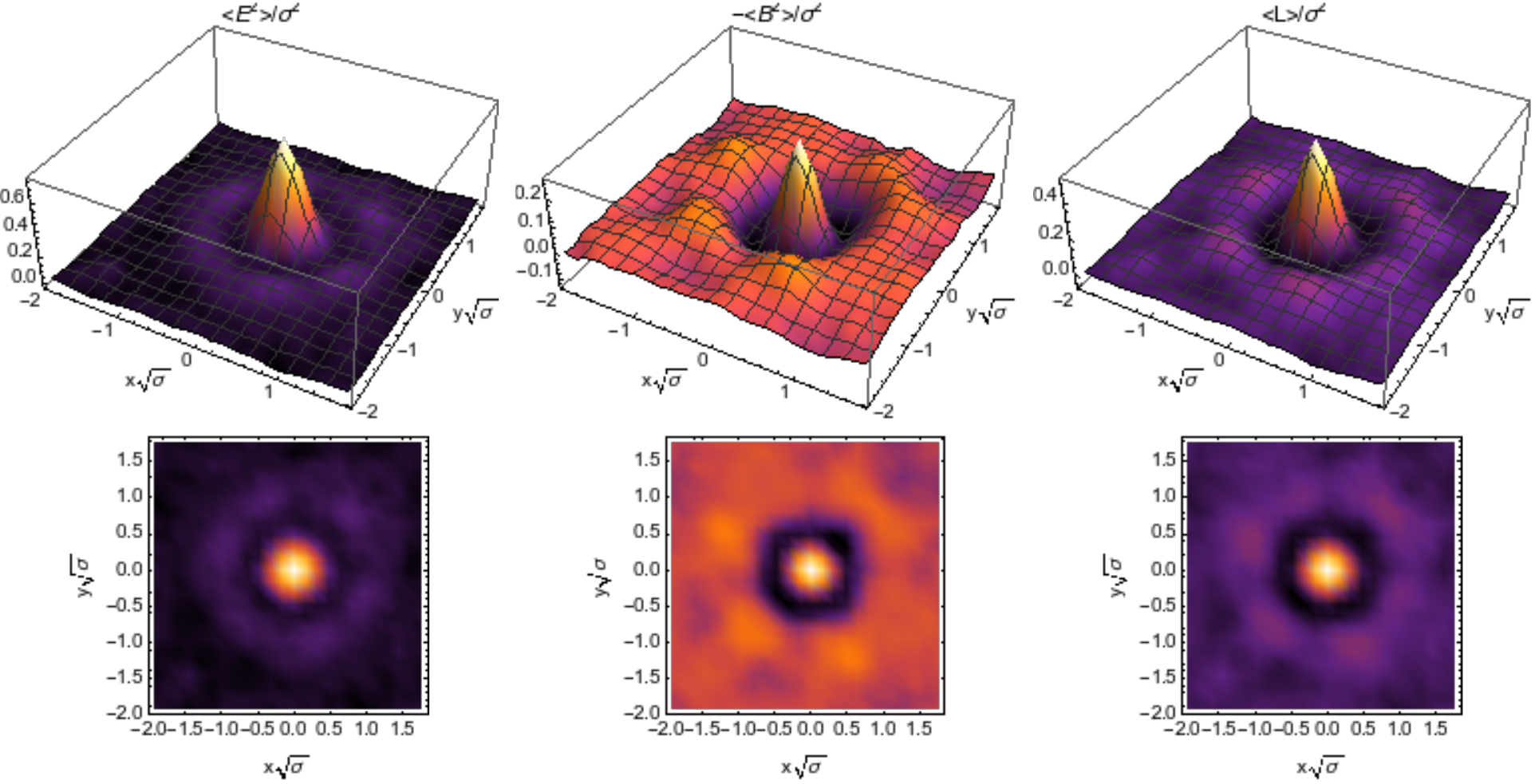}
\includegraphics[width=\LagFigSize]{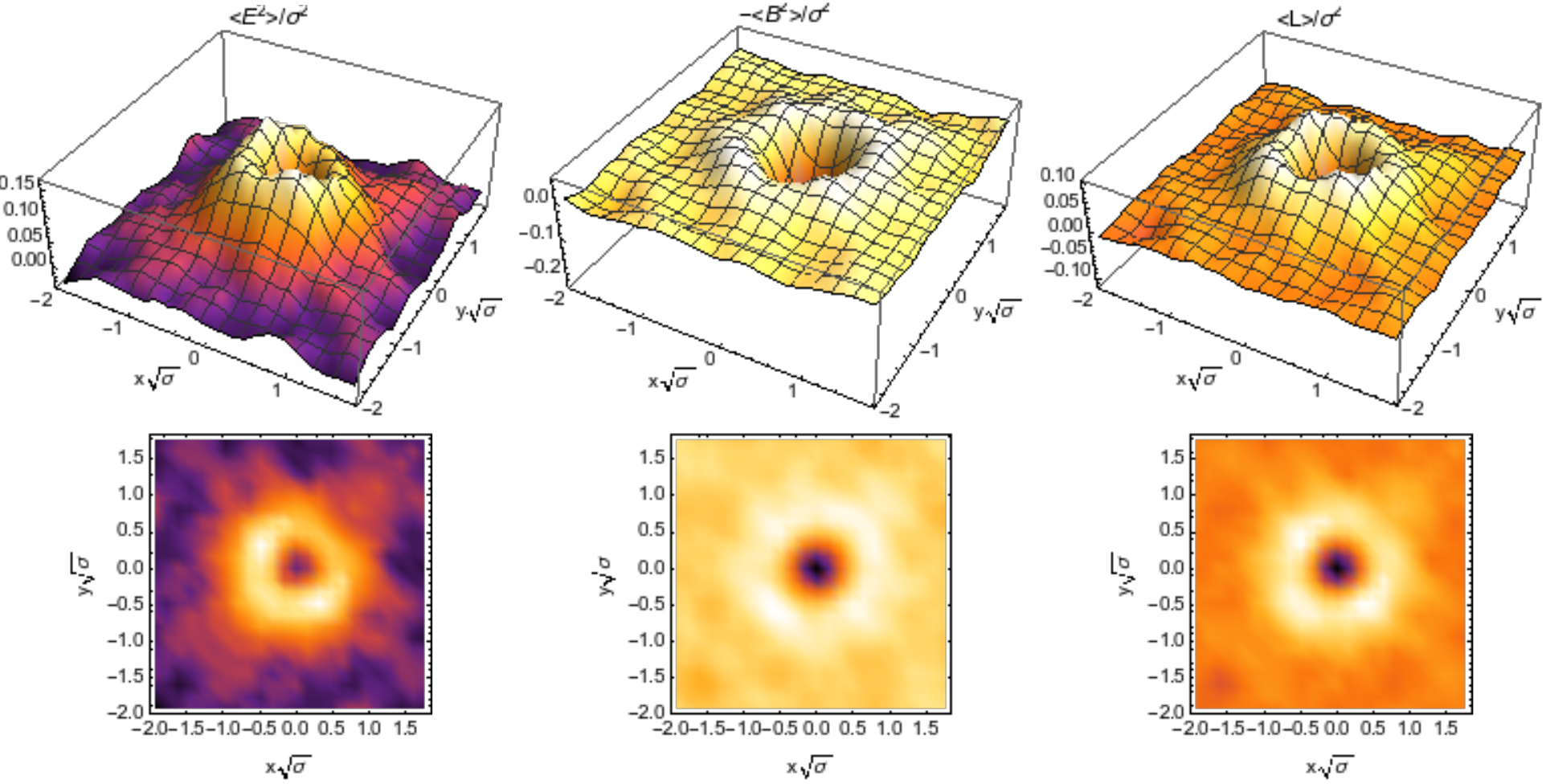}
\par\end{centering}
\caption{3D plots of the Lagrangian $\cal L$, $E^2$ and $B^2$ field densities in the mediator plane, for an inter-charge distance $R=10 a= 1.610 / \sqrt \sigma $. We show the excited states ${\Sigma_g^+}^{*}$ (top), ${\Pi_u}^*$   (bottom), whereas the groundstates are shown in Fig. \ref{fig:3dplot_L_xy}. The distance and energy are shown in string tension units $\sqrt \sigma$. The magnitude of the densities is the same of Fig. \ref{fig:field_L_xy}.
\label{fig:3dplot_PIu_xy}}
\end{figure}

A departure from the most common profile of the flux tubes, in general dominated by a Gaussian or exponential-like profile is indeed observed in the channels $\Pi_u^{}$ and $\Pi_u^{*}$, as shown in Figs. \ref{fig:3dplot_L_xy} and \ref{fig:3dplot_PIu_xy}.

For these cases, there is a clear difference in the magnetic field component, which squared field density $- \langle B^2 \rangle$ is negative in the median point, as we show in Fig. \ref{fig:field_B_xy}.

Notice this is again in contradistinction with a transverse standing p-wave, which should vanish in the origin.
The only way for an angular momentum $\Lambda= 1$ not to vanish in the origin is to have a particle with a spin, since the wavefunction with a finite orbital angular momentum should vanish in the $z$ axis, as illustrated in Fig. \ref{fig:quantumnumbers}.

Thus our results suggest, among the possible components of the $\Pi_u^{}$ and $\Pi_u^{*}$ states, the component with a dynamical gluon is dominant. These states are essentially hybrid states, with a static quark, a static antiquark and a dynamical gluon.

%

\section{Conclusions and outlook \label{sec:conclu}}

We succeed in showing the techniques of Ref. \cite{Cardoso:2013lla} to study the field densities can be extended to excited flux tubes. We improve the work of Ref. \cite{Bicudo:2018yhk} and clarify a discrepancy it had with Ref.  \cite{Mueller:2018fkg}.

We compute the potentials and flux tube densities for several excitations of the pure $SU(3)$ flux tubes produced by two static $3$ and $\bar 3$ sources.
We consider radial excitations of the groundstate $\Sigma^+_g$, the first axial parity excitation $\Sigma^+_u$ and the first angular excitation $\Pi_u$. 
We select the main excited states, up to three states, in each quantum number. 

In our results, Figs.  \ref{fig:field_L_z}, \ref{fig:field_L_xy}, \ref{fig:field_E_z}, \ref{fig:field_B_z}, \ref{fig:field_E_xy} and \ref{fig:field_B_xy} , we compare the chromoelectric and the chromomagnetic field densities, both in the mediator plane and in the charge axis. We analyse several aspects of the flux tubes as well in our 3D Figs. \ref{fig:3dplot_L_xy} and \ref{fig:3dplot_PIu_xy}, comparing with the bosonic Nambu-Goto string model. 

In particular we find evidences the flux tube cannot be described by a string model with transverse modes only, and we also find evidence for hybrid $\Pi_u^{}$ and $\Pi_u^{*}$ states, where an explicit gluon is coupled to the flux tube. 

As an outlook, we plan to continue this first study of SU(3) flux tubes, when we will be able to use much more computational power. 
It is important to be able to compute flux tubes for a larger operator basis and more quantum numbers. It will also be interesting to further clarify the questions raised by our results, as analysed in Section \ref{sec:analysis}. 
Also notice the square of the chromoelectric or chromomagnetic fields and the Lagrangian densities operators suffer from ultraviolet divergences in the lattice gauge field theories, therefore the absolute magnitude of their expectation values should depend on the lattice spacing $a$. 
It is also important to extrapolate to the infinite volume limit.
Since all these studies will require computational power beyond our resources, we leave it for the future.

\vspace{0.25cm}
\textbf{Acknowledgments}
\vspace{0.1cm}

Pedro Bicudo is very thankful to Bastian Brandt, Richard Brower, Lasse M\"uller and Marc Wagner for discussions on flux tubes. 
Nuno Cardoso is supported by FCT under the contracts SFRH/BPD/109443/2015.
We also acknowledge the use of CPU and GPU servers of PtQCD, partly supported by NVIDIA, CeFEMA and the FCT grant 
UID/CTM/04540/2013.

\bibliographystyle{elsarticle-num}
\bibliography{excited}{}

\end{document}